\documentclass[aps,prd,twocolumn,showpacs,amsmath,amssymb]{revtex4-1}
\usepackage{amsmath}
\usepackage{amsmath}
\usepackage{graphicx}
\usepackage{subfigure}
\usepackage{epstopdf}
\usepackage{color}
\usepackage{multirow}
\usepackage{setspace}
\usepackage{overpic}
\usepackage{amssymb}
\usepackage[bookmarksnumbered, pdfstartview=FitH,colorlinks,urlcolor=blue, citecolor=blue,linkcolor=blue] {hyperref}
\usepackage{lineno}
\usepackage{bm}
\usepackage{rotating}
\usepackage{siunitx}
\usepackage[utf8]{inputenc}

\let\oldequation\equation
\let\oldendequation\endequation

\renewenvironment{equation}
  {\linenomathNonumbers\oldequation}
  {\oldendequation\endlinenomath}

\begin{document}

\title{\bf \boldmath
Measurements of the branching fractions of the inclusive decays $D^{0}(D^{+})\to \pi^+\pi^+\pi^-X$}

\author{
M.~Ablikim$^{1}$, M.~N.~Achasov$^{13,b}$, P.~Adlarson$^{73}$, R.~Aliberti$^{34}$, A.~Amoroso$^{72A,72C}$, M.~R.~An$^{38}$, Q.~An$^{69,56}$, Y.~Bai$^{55}$, O.~Bakina$^{35}$, I.~Balossino$^{29A}$
% ...
\\
\vspace{0.2cm}
(BESIII Collaboration)\\
\vspace{0.2cm} {\it
$^{1}$ Institute of High Energy Physics, Beijing 100049, People's Republic of China\\
% ...
$^{13}$ G.I. Budker Institute of Nuclear Physics SB RAS (BINP), Novosibirsk 630090, Russia\\
% ...
$^{29}$ INFN Sezione di Ferrara, (A)INFN Sezione di Ferrara, I-44122, Ferrara, Italy; (B)University of Ferrara, I-44122, Ferrara, Italy\\
% ...
$^{34}$ Johannes Gutenberg University of Mainz, Johann-Joachim-Becher-Weg 45, D-55099 Mainz, Germany\\
$^{35}$ Joint Institute for Nuclear Research, 141980 Dubna, Moscow region, Russia\\
% ...
$^{38}$ Liaoning Normal University, Dalian 116029, People's Republic of China\\
% ...
$^{55}$ Southeast University, Nanjing 211100, People's Republic of China\\
$^{56}$ State Key Laboratory of Particle Detection and Electronics, Beijing 100049, Hefei 230026, People's Republic of China\\
% ...
$^{69}$ University of Science and Technology of China, Hefei 230026, People's Republic of China\\
% ...
$^{72}$ University of Turin and INFN, (A)University of Turin, I-10125, Turin, Italy; (B)University of Eastern Piedmont, I-15121, Alessandria, Italy; (C)INFN, I-10125, Turin, Italy\\
$^{73}$ Uppsala University, Box 516, SE-75120 Uppsala, Sweden\\
% ...
$^{b}$ Also at the Novosibirsk State University, Novosibirsk, 630090, Russia\\
% ...
}
\vspace{0.4cm}
}

\begin{abstract}
Using $e^+e^-$ annihilation data corresponding to an integrated
luminosity of 2.93 fb$^{-1}$ taken at a
center-of-mass energy of 3.773 GeV with the BESIII detector, we report
the first measurements of the branching fractions of the inclusive
decays $D^0\to \pi^+\pi^+\pi^-X$ and $D^+\to \pi^+\pi^+\pi^-X$, where
pions from $K^0_S$ decays have been excluded from the
$\pi^+\pi^+\pi^-$ system
and $X$ denotes any possible particle combination.  The branching
fractions of $D^0(D^+)\to \pi^+\pi^+\pi^-X$ are determined to be
${\mathcal B}(D^0\to \pi^+\pi^+\pi^-X)=(17.60\pm0.11\pm0.22)\%$ and
${\mathcal B}(D^+\to \pi^+\pi^+\pi^-X)=(15.25\pm0.09\pm0.18)\%$, where
the first uncertainties are statistical and the second systematic.
\end{abstract}

\maketitle
\section{Introduction}

In recent years, tests of lepton flavor universality (LFU) have become
a very hot topic in heavy flavor physics.  The world averages of the
ratios $R(D)=\mathcal {B}(B\to D\tau\nu_{\tau})/\mathcal {B}(B\to
D\ell\nu_{\ell})$ and $R(D^*)=\mathcal {B}(B\to
D^*\tau\nu_{\tau})/\mathcal {B}(B\to D^*\ell\nu_{\ell})$, with $\ell =
e$ or $\mu$, deviate from the Standard Model (SM) predictions by more
than $1.4\sigma$ and $2.8\sigma$, respectively~\cite{LFUtest}.
Additionally, the LHCb experiment reported the ratio of branching
fractions, $R(D^{*-})=\mathcal {B}(B^0\to
D^{*-}\tau^+\nu_{\tau})/\mathcal {B}(B^0\to D^{*-}\mu^+\nu_{\mu})$,
based on 3 fb$^{-1}$ of $pp$ data taken at 7 and 8 TeV (Run
I)~\cite{LHCb1,LHCb2}, which had the smallest statistical uncertainty
at the time and was consistent with the SM prediction within $1.1\sigma$.
However, the LHCb measurement is limited by the knowledge of the
normalization channel $\mathcal {B}(B^0\to
D^{*-}\pi^+\pi^+\pi^-)$.
Future data taken at the Belle II and LHCb experiments will help to
further improve the accuracy of the branching fractions and tests of
LFU.

In these tests, the analyses adopt the decay chain of $B^0\to
D^{*-}\tau^+\nu_{\tau}$ with $\tau^+ \to \pi^+\pi^+\pi^-\bar
\nu_\tau$, where the leading and sub-leading background sources are from $B^{0,+}\to D_s^+ + anything$ with $D^+_s\to \pi^+\pi^+\pi^-X$  and $B^{0,+}\to D^0(D^+)+anything$ with $D^{0}(D^{+})\to
\pi^+\pi^+\pi^-X$ (where $\pi^\pm$s from $K^0_S$ decays have been excluded
from $\pi^+\pi^+\pi^-$ and $X$ denotes any possible particle
combination), respectively. Unfortunately, information on inclusive decays of
charmed mesons into final states containing $\pi^+\pi^+\pi^-$ is
sparse. Measurements of the full and partial decay branching fractions
of the inclusive decays $D^+_s\to \pi^+\pi^+\pi^-X$ and
$D^{0}(D^{+})\to \pi^+\pi^+\pi^-X$ offer important inputs to precisely
test LFU with semileptonic $B$ decays.

Recently, the BESIII Collaboration reported the first measurement of
the branching fraction of the inclusive decay $D^+_s \to
\pi^+\pi^+\pi^-X$~\cite{bes3-Ds-3pix}.  The branching
fraction obtained is greater than the sum of the branching fractions of the
known exclusive $D^+_s$ decays containing $\pi^+\pi^+\pi^-$ by around
25\%, thereby implying that many exclusive $D^+_s$ decays containing
$\pi^+\pi^+\pi^-$ are still unmeasured.  The sums of the branching
fractions of the known exclusive $D^0$ and $D^+$ decays containing
$\pi^+\pi^+\pi^-$~\cite{pdg2020,wuxiao,kaikai,lanxing}, as summarized
in Appendix~\ref{sec:app}, are $(16.05\pm 0.47)$\% and $(14.74\pm
0.53)$\%, respectively.  The measurements of the branching fractions
of the inclusive decays $D^{0}(D^+)$ can offer a check on the known
exclusive $D^{0}(D^+)$ decays containing $\pi^+\pi^+\pi^-$. A
measurable difference between the branching fractions of inclusive and
exclusive decays would indicate that either some exclusive decays are
not measured or that some known exclusive decays are overestimated.

In this paper, we report the first measurements of the branching fractions of $D^0 \to \pi^+\pi^+\pi^-X$ and $D^+ \to \pi^+\pi^+\pi^-X$
by analyzing $2.93\ \text{fb}^{-1}$ of $e^+e^-$ collision data~\cite{lum_bes3} taken at a center-of-mass energy of 3.773 GeV with the BESIII detector. Throughout this paper, charge conjugate decays are always implied.

\section{BESIII detector and Monte Carlo simulation}

The BESIII detector is a magnetic
spectrometer~\cite{BESIII} located at the Beijing Electron
Positron Collider (BEPCII)~\cite{Yu:IPAC2016-TUYA01}. The
cylindrical core of the BESIII detector consists of a helium-based
 multilayer drift chamber (MDC), a plastic scintillator time-of-flight
system (TOF), and a CsI (Tl) electromagnetic calorimeter (EMC),
which are all enclosed in a superconducting solenoidal magnet
providing a 1.0~T magnetic field~\cite{dectector}. The solenoid is supported by an
octagonal flux-return yoke with resistive plate counter muon
identifier modules interleaved with steel. The solid angle coverage for detecting charged particles is 93\% over $4\pi$. The
charged-particle momentum resolution at $1~{\rm GeV}/c$ is
$0.5\%$, and the resolution of the specific ionization energy loss~(d$E$/d$x$) is $6\%$ for the electrons
from Bhabha scattering. The EMC measures photon energies with a
resolution of $2.5\%$ ($5\%$) at $1$~GeV in the barrel (end cap)
region. The time resolution of the TOF barrel part is 68~ps, while
that of the end cap part is 110~ps.
More details about the design and performance of the BESIII detector are given in Ref.~\cite{BESIII}.

Simulated samples produced with {\sc geant4}-based~\cite{geant4}
Monte Carlo (MC) software, which includes the geometric description of
the BESIII detector and the detector response, are used to determine
the detection efficiency and to estimate background contributions. The
simulations include the beam energy spread and initial state radiation
in the $e^+e^-$ annihilations modeled with the generator {\sc
  kkmc}~\cite{kkmc}.  The inclusive MC samples consist of the
production of $D\bar{D}$ pairs with quantum coherence (QC) for neutral
$D$ modes, the non-$D\bar{D}$ decays of the $\psi(3770)$, the initial
state radiation production of the $J/\psi$ and $\psi(3686)$ states,
and the continuum processes.  The known decay modes are modeled with
{\sc evtgen}~\cite{evtgen} using the branching fractions taken from
the PDG~\cite{pdg2020}, and the remaining unknown decays of the
charmonium states are modeled by {\sc
  lundcharm}~\cite{lundcharm,lundcharm2}. Final state radiation is
incorporated using {\sc photos}~\cite{photos}.

\section{Method}
\label{sec:method}

As the peak of the $\psi(3770)$ resonance lies just above the $D\bar
D$ threshold, it decays predominately into $D\bar D$ meson pairs. We
take advantage of this by using a double-tag (DT) method, which was
first developed by the MARKIII Collaboration~\cite{MARKIII1,MARKIII2}
to determine absolute branching fractions. The single-tag (ST) $\bar
D^0$($D^-$) mesons are selected by using the two hadronic decay modes
$\bar D^0\to K^+\pi^-$ and $D^-\to K^+\pi^-\pi^-$, respectively, which
have a relatively large branching fraction and low background.

Events where $D^{0}(D^{+})$ decaying into signal particles can be
selected in the presence of ST $\bar D^0$($D^-$) mesons are called DT
events. To compensate for the differences of the $3\pi$ invariant mass
$M_{3\pi}$ distributions between data and MC simulation and to
consider the signal migration among different ($M_{3\pi}$) intervals, we
determine the partial branching fractions of the $D^{0}(D^{+})\to
\pi^+\pi^+\pi^-X$ decays in bins of $M_{3\pi}$ at the production
level. The number of produced DT events and the numbers of observed DT
events are related in bins through a detector response matrix that
accounts for detector efficiency and detector resolution,
\begin{equation}
N_{\rm obs}^i = {\sum_{j=1}^{N_{\rm intervals}} \epsilon_{ij}N^j_{\rm prod}},
\label{eq:DTyield}
\end{equation}
where $N^i_{\rm obs}$ is the number of signal events observed in the $i$-th $M_{3\pi}$ interval,
$N^j_{\rm prod}$ is the number of signal events produced in the $j$-th $M_{3\pi}$ interval,
and $\epsilon_{ij}$ is the efficiency matrix describing the detection efficiency and migration effect across each $M_{3\pi}$ interval. The statistical uncertainties of $\epsilon_{ij}$ due to the limited size of the signal MC simulation sample are considered as a source of systematic uncertainties, as discussed in \ref{SYS}.

The number of the inclusive decays $D^0(D^+)\to \pi^+ \pi^+ \pi^- X$ produced in the $i$-th $M_{3\pi}$ interval is obtained by solving Eq.~(\ref{eq:DTyield}) for $N^i_{\rm prod}$, which gives
\begin{equation}
N_{\rm prod}^i = {\sum_{j=1}^{N_{\rm intervals}} (\epsilon^{-1})_{ij}N^j_{\rm obs}}.
\label{Eq11}
\end{equation}
The statistical uncertainty of $N^i_{\rm prod}$ is given by
\begin{equation}
(\sigma_{\rm stat}(N_{\rm prod}^i))^2 = {\sum_{j=1}^{N_{\rm intervals}} (\epsilon^{-1})^2_{ij}(\sigma_{\rm stat}(N^j_{\rm obs}))^2},
\end{equation}
where $\sigma_{\rm stat}(N_{\rm obs}^j)$ is the statistical uncertainty of $N^j_{\rm obs}$.

The partial branching fraction of the $i$-th $M_{3\pi}$ interval is determined by
\begin{equation}
{\rm{d}\mathcal B_{\rm sig}}= \frac{N_{\rm prod}^i}{N_{\rm ST}/\epsilon_{\rm tag}},
\label{branch}
\end{equation}
where $N_{\rm ST}/\epsilon_{\rm tag}$ is the efficiency corrected yield of the ST $\bar{D}^0(D^-)$ mesons. The partial branching fractions are summed to obtain the total branching fraction $\mathcal B_{\rm sig}$.

Since the measurement of the branching fraction of $D^0\to
\pi^+\pi^+\pi^-X$ is affected by quantum coherence (QC) in the $D^0
\bar D^0$ system,
the branching fraction of $D^0\to \pi^+\pi^+\pi^-X$ measured with the
tag mode $\bar D^0\to K^+\pi^-$ needs to be
corrected by
\begin{equation}\label{equ:QCcorrect}
{\rm d\mathcal B}^{\rm corr}_{\rm sig} = f^{\rm corr}_{\rm QC}\times {\rm d\mathcal B}_{\rm sig},
\end{equation}
where
\begin{equation}
\label{fcorr}
f^{\rm corr}_{\rm QC}=\frac{1}{1-C_f(2f_{CP+}-1)},
\end{equation}
${\mathcal B}_{\rm sig}$ is the branching fraction to be measured, and
$C_{f}$ denotes the strong-phase factor calculated by
\begin{equation}
\label{corr}
C_{f} = \frac{2rR\cos\delta}{1 + r^2}.
\end{equation}
In Eq.~(\ref{corr}), $r$ is the ratio between the
doubly-Cabibbo-suppressed and Cabibbo-favored amplitudes for $D\to
K^\pm\pi^\mp$, $\delta$ is the strong phase difference between the two
amplitudes and $R=1$ is the coherence factor for $D\to
K^\pm\pi^\mp$~\cite{refcp3, refcp4}.  Table~\ref{CP_parameters}
summarizes the parameters of $r$ and $\delta$ for $D\to K^\pm\pi^\mp$,
which give $C_f=(-11.3^{+0.4}_{-0.9})\%$ for $D\to K^\pm\pi^\mp$.

\begin{table}[htbp]
\centering
\caption{
Input parameters for the QC correction.}
\label{CP_parameters}

\begin{tabular}{cc}
  \hline\hline
Parameter                      &   Value                                 \\
  \hline
$r_{K\pi}$                           &    0.0586$\pm$0.0002~\cite{A1}                      \\
$\delta_{K\pi}$                      &    $(194.7_{-17.0}^{+8.4})^\circ$~\cite{A1}        \\
\hline\hline
\end{tabular}
\end{table}

In Eq.~\eqref{fcorr}, $f_{CP+}$ is the fraction of the $CP$-even ($+$) component. According to Refs.~\cite{refcp1,refcp2},
$f_{CP+}$ is calculated by
\begin{equation}
f_{CP+} = \frac{N^{+}}{N^++N^-}
\nonumber
\end{equation}
with
\begin{equation}
N^{\pm} = \frac{M^{\pm}_{\rm measured}}{S^{\pm}},
\nonumber
\end{equation}
\begin{equation}
S_{\pm} = \frac{S^{\pm}_{\rm measured}}{1 - \eta_{\pm} y_D}.
\nonumber
\end{equation}
Here, $N^{\pm}$ is the ratio of DT and ST yields with the $CP$ even
and odd $CP{\mp}$ tags, $M^{\pm}_{\rm measured}$ denotes the
number of DT candidates for a signal channel versus $CP{\mp}$ tags,
and $S^{\pm}_{\rm measured}$ is the number of ST candidates for
$CP{\pm}$ decay modes.  Finally, $\eta_{\pm}=\pm1$ for $CP\pm$ mode
and $y_D= (0.62\pm0.08)$\% is the mixing parameter of $D^0\bar D^0$
taken from the latest average by the PDG~\cite{pdg2020}.

\section{The ST analysis}

The charged kaons and pions are selected and identified with the same
criteria as in Refs.~\cite{bes3-pimuv,bes3-etaX}.  For each charged
track, the polar angle ($\theta$) is required to be within the MDC
acceptance $|\rm{cos(\theta)}|<0.93$, where $\theta$ is defined with
respect to the $z$ axis, which is the symmetry axis of the MDC.  The
distance of the charged track's closest approach relative to the
interaction point is required to be within $10$~cm along the $z$ axis
and within $1$~cm in the plane perpendicular to the $z$ axis.
Particle identification (PID) for charged tracks combines the
measurements of d$E$/d$x$ in the MDC and the flight time in the
TOF to form probabilities $\mathcal{L}(h) (h = K, \pi)$ for each
hadron $(h)$ hypothesis. Charged tracks are assigned as kaons or
pions if their probabilities satisfy one of the two hypotheses,
$\mathcal{L}(K) > \mathcal{L}(\pi)$ or $\mathcal{L}(\pi) > L(K)$,
respectively.  In the selection of $\bar D^0\to K^+\pi^-$ candidates,
background contributions from cosmic rays and Bhabha events are
rejected with the following requirements. First, the two charged
tracks must have a TOF time difference less than 5 ns and they must
not be consistent with being a muon pair or an electron–positron
pair. Second, there must be at least one EMC shower with an energy
greater than 50 MeV or at least one additional charged track detected
in the MDC.

To distinguish the ST $\bar D$ mesons from combinatorial background, we define the two kinematic variables of energy difference $\Delta E$
and the beam-constrained mass $M_{\rm BC}$ as
\begin{equation}
\Delta E \equiv E_{\bar D} - E_{\rm beam},
\label{eq:deltaE}
\end{equation}
and
\begin{equation}
M_{\rm BC} \equiv \sqrt{E^{2}_{\rm beam}/c^4-|\vec{p}_{\bar D}|^{2}/c^2}.
\label{eq:mBC}
\end{equation}
Here, $E_{\rm beam}$ is the beam energy, and
$E_{\bar D}$ and $\vec{p}_{\bar D}$  are the energy and momentum of
the $\bar D$ candidate in the rest frame of the $e^+e^-$ system.
For each ST mode, if there are multiple candidates in an event,
the one with the smallest $|\Delta E|$ is kept.
The ST $\bar D^0$ and $D^-$ candidates are required to satisfy
$|\Delta E|<25$\,MeV and $|\Delta E|<20$\,MeV, respectively, which corresponds to about $\pm 3\sigma$ of the fitted resolution.

\begin{figure}[htbp]
\centering
\includegraphics[width=1.0\linewidth]{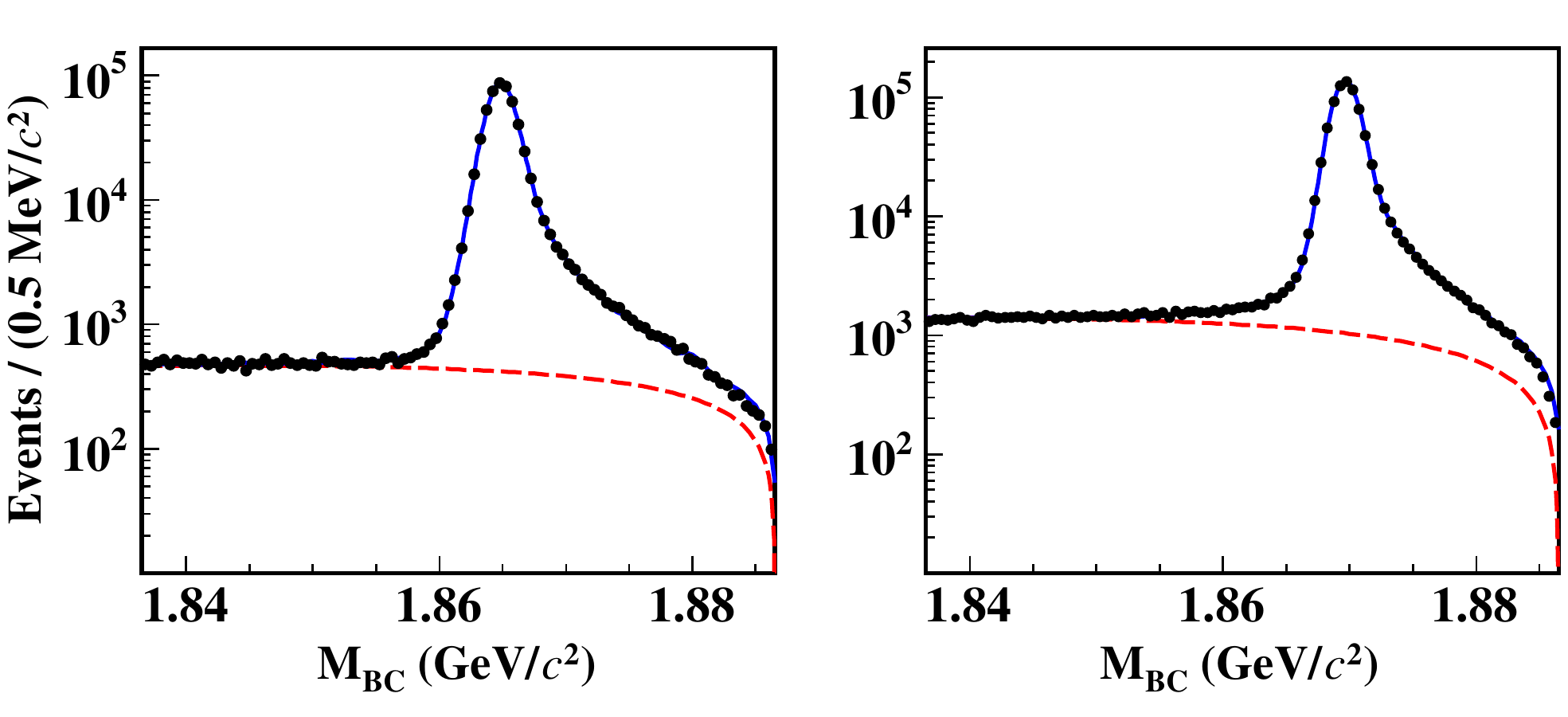}
\caption{Fits to the $M_{\rm BC}$ distributions of the ST candidates for  $\bar D^0\to K^+\pi^-$ (left) and $D^-\to K^+\pi^-\pi^-$ (right), where the points with error bars are data, the blue solid curves are the fit results, and the red dashed curves are the fitted combinatorial backgrounds.}
\label{fig::datafit_MassBC}
\end{figure}

To determine the yields of ST $\bar D^0$ and $D^-$ mesons, maximum
likelihood fits are performed on the corresponding $M_{\rm BC}$
distributions of the accepted ST candidates.  In the fits, the signal
shape of $\bar D^0$ or $D^-$ is modeled by an MC-simulated shape
convolved with a double-Gaussian function, which is a sum of two
Gaussian functions with free parameters, describing the resolution
difference between data and MC simulation due to two asymmetrical
tails of signal.  The combinatorial background shape is described by
an ARGUS function~\cite{ARGUS}.  The resultant fits to the $M_{\rm
  BC}$ distributions are shown in Fig.~\ref{fig::datafit_MassBC}.  The
yields of ST $\bar D^0$ and $D^-$mesons are $548031\pm775$ and
$812109\pm1896$, respectively, where the uncertainties are statistical
only.  The efficiencies of reconstructing the ST $\bar D$ mesons are
estimated to be $(67.70\pm0.08)\%$ and $(51.58\pm0.04)\%$ for neutral
$\bar D^0$ decay and charged $D^-$ decay, respectively, by analyzing
the inclusive MC sample with the same procedure as that for data.

\section{The DT analysis}
\subsection{Selection of $D^{0}(D^{+})\to \pi^+\pi^+\pi^-X$}

The candidates for $D^{0}(D^{+})\to \pi^+\pi^+\pi^-X$ are selected in
the presence of the ST $\bar D$ mesons.  We require that there are at
least three charged pions which have not been used in the ST
selection. If there are more than one $\pi^-$ or two $\pi^+$ mesons
reconstructed on the signal side, only the one with the
fastest two $\pi^+$s and the fastest $\pi^-$ are kept for further analysis.

To reject background components from $D^{0}(D^+)\to \pi^+ K_S^0(\to
\pi^+\pi^-)X$ decays ($K_S^0$ BKG1), the invariant mass of any
$\pi^+\pi^-$ combination from the three selected pions is required to
fulfill $|M_{\pi^+\pi^-}-0.4977|>0.030~$~GeV/$c^2$.  Here, there is no decay length requirement for $K_S^0$ BKG1. In addition,
another $K_S^0$ background ($K_S^0$ BKG2), where one of the $\pi$
mesons comes from the chosen three pions and another assumed $\pi$
meson comes from a remaining oppositely charged track without PID, is
rejected. The $K^0_S$ candidate is selected through the following
selection criteria: First, the $\pi^+\pi^-$ pair is constrained to
originate from a common vertex. Second, the invariant mass of the
$\pi^+\pi^-$ pair is in the range of
$|M_{\pi^+\pi^-}-0.4977|>0.012~$~GeV/$c^2$. Third, the decay length of
$K_S^0$ candidates is greater than two standard deviations of the
vertex resolution away from the interaction point.

To further suppress the remaining background contributions of $K_S^0$
BKG2 and $D^0\to \pi^+\pi^- K^0_S(\to \pi^+\pi^-)$ ($K_S^0$ BKG3), the
recoil masses of the $\bar D^0\pi^+\pi^-$ combinations are required to
be $|M_{\pi^+\pi^-}-0.4977|>0.080~$~GeV/$c^2$.  For background
contributions from $D^0\to \pi^+\pi^-\pi^0 K_{S}^{0}$ ($K_S^0$ BKG3),
a similar requirement is applied on $\bar D^0\pi^+\pi^-\pi^0$
combinations if a good $\pi^0$ is found.  The $\pi^0$ candidates are
reconstructed via the $\pi^0\to \gamma\gamma$ decay and the opening
angle between the photon candidate and the nearest charged track is
required to be greater than $10^\circ$. Any photon pair with an
invariant mass between $(0.115, 0.150)$ GeV$/c^2$ is regarded as a
$\pi^0$ candidate, and a kinematic fit is imposed on the photon pair
to constrain its invariant mass to the known $\pi^0$
mass~\cite{pdg2020}.

\begin{figure}[htbp]
\centering
\includegraphics[width=1.0\linewidth]{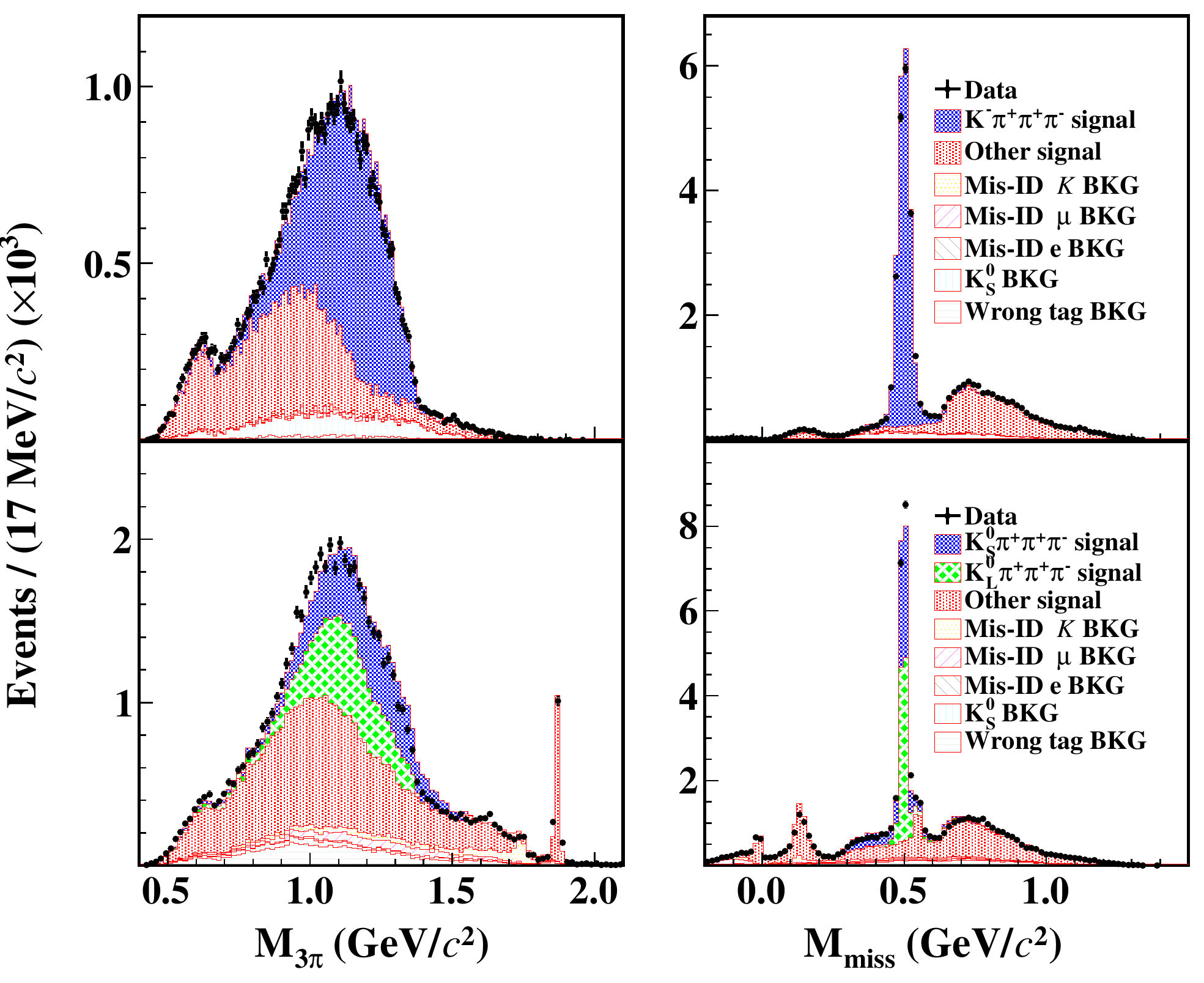}
\caption{Comparisons of the $M_{3\pi}$ (left) and $M_{\rm miss}$
  (right) distributions of the DT candidates for $D^0\to
  \pi^+\pi^+\pi^-X$ (top) and $D^+\to \pi^+\pi^+\pi^-X$ (bottom),
  where the points with error bars are data and the color filled
  histograms are the inclusive MC sample. To ensure a $\bar D$ in the
  ST side, events must satisfy the requirements mentioned in
  the text and an additional requirement of $|M_{\rm BC}-M_D|<0.005$
  GeV/$c^2$, where $M_D$ is the known $D$ mass~\cite{pdg2020}.
\label{fig::m3pi}
}
\end{figure}

\begin{figure}[htbp]
\centering
\includegraphics[width=1.0\linewidth]{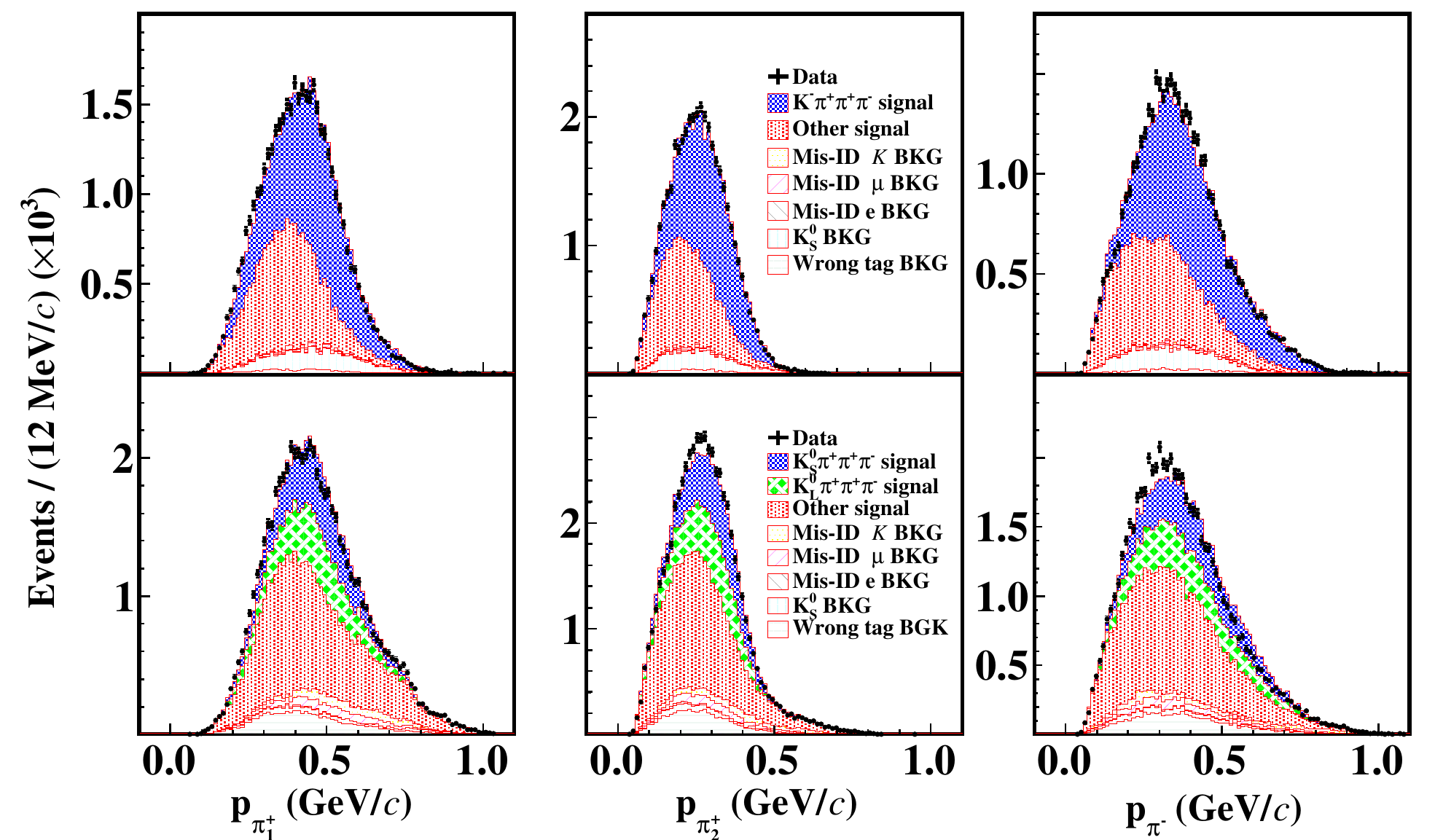}
\caption{Comparisons of the momentum distributions of the selected
  pions of the DT candidates for $D^0\to \pi^+\pi^+\pi^-X$ (top) and
  $D^+\to \pi^+\pi^+\pi^-X$ (bottom), where the points with error bars
  are data and the color filled histograms are the inclusive MC
  sample. To ensure a $\bar D$ in the ST side, events must satisfy the
  requirements mentioned in the text and an additional requirement of
  $|M_{\rm BC}-M_D|<0.005$ GeV/$c^2$, where $M_D$ is the known $D$
  mass~\cite{pdg2020}, and $\pi^+_1$ and $\pi^+_2$ denote the higher
  and lower momentum $\pi^+$s, respectively.
\label{fig::momentum}}
\end{figure}

Figure \ref{fig::m3pi} shows the comparison
of the distributions of $M_{3\pi}$ and $M_{\rm miss}$ of the selected charged pions for the accepted DT candidates between data and the inclusive MC sample. Throughout this paper, $M_{3\pi}$ is the invariant mass of the selected $\pi^+\pi^+\pi^-$ combination and
$M_{\rm miss}$ is the missing mass of the $\bar D \pi^+\pi^+\pi^-$ combination given by
\begin{equation}
M_{\rm miss}^2=(2E_{\rm beam}-E_{\bar{D}}-E_{3\pi})^2/c^4-\left|-\vec{p}_{\bar {D}}-\vec{p}_{3\pi}\right|^2/c^2,
\end{equation}
where $E_{3\pi}$ and $\vec{p}_{3\pi}$ are the total energy and momentum of the selected $\pi^+\pi^+\pi^-$ combination of the signal side in the $e^+e^-$ center-of-mass frame.
Small inconsistencies between data and the inclusive MC sample around $(0.9, 1.2)$~GeV/$c^2$ in the $M_{3\pi}$ distributions are mainly
due to imperfect simulations of multi-body hadronic decays with low branching fractions.
For the $D^0\to \pi^+\pi^+\pi^-X$ and $D^+\to \pi^+\pi^+\pi^-X$ decays, the largest signal components are from
$D^0\to K^-\pi^+\pi^+\pi^-$ and $D^+\to \bar K^0\pi^+\pi^+\pi^-$ decays, respectively, and they form peaks
around the known $\bar K$ mass in the $M_{\rm miss}$ distributions as expected.
For $D^+\to \pi^+\pi^+\pi^-X$, the peak around the known $D^+$ mass in the $M_{3\pi}$ distribution is from
$D^+\to \pi^+\pi^+\pi^-$;
the peaks around zero and 0.135 GeV/$c^2$ in the $M_{\rm miss}$ distribution are from
$D^+\to \pi^+\pi^+\pi^-$ and $D^+\to \pi^+\pi^+\pi^-\pi^0$, respectively;
the peak around 0.51 GeV/$c^2$ in the $M_{\rm miss}$ distribution is mainly from $D^+\to \pi^+\pi^+\pi^-\eta$. Comparisons of the distributions of momenta of the selected three charged pions are in good agreement, as shown in Fig. \ref{fig::momentum}.

Background analysis based on the inclusive MC sample shows that there
are still some remaining background components, even after imposing
all the aforementioned background rejection requirements.  One is from
events with wrongly tagged $\bar D$ decays ($i.e.$ ST decays are not
$\bar D^0\to K^+\pi^-$ or $D^-\to K^+\pi^-\pi^-$) and the non-$D\bar
D$ process, labeled as ``wrong tag''.  Another background component
results from events with correctly tagged $\bar D$ decays, but
incorporating particle mis-identifications of $K\to \pi$ (Mis-ID $K$
BKG), $\mu\to \pi$ (Mis-ID $\mu$ BKG) and $e\to \pi$ (Mis-ID $e$ BKG)
as well as the remaining $K_S^0$ background ($K_S^0$ BKG).  The
background components are also shown in Figs.~\ref{fig::m3pi}
and \ref{fig::momentum}.

\subsection{DT signal yields}

To minimize a possible efficiency dependence of various $D$ decay
modes and offer finer information for the LFU tests in the
semileptonic $B$ decays, the partial branching fractions of
$D^0 \to \pi^+\pi^+\pi^-X$ and $D^+ \to \pi^+\pi^+\pi^-X$ are measured in nine
and ten $M_{3\pi}$ bins, respectively. For $D^0\to \pi^+\pi^+\pi^-X$
and $D^+\to \pi^+\pi^+\pi^-X$, the lower boundaries of the intervals
are chosen as [0.40, 0.55, 0.70, 0.85, 1.00, 1.15, 1.30, 1.45, 1.60,
  1.75]~GeV/$c^2$ and [0.40, 0.55, 0.70, 0.85, 1.00, 1.15, 1.30, 1.45,
  1.60, 1.75, 2.00]~GeV/$c^2$, respectively.  For $D^+\to
\pi^+\pi^+\pi^-X$, the interval [1.75, 2.00] GeV/$c^2$ is added to
specifically consider the hadronic decay $D^+\to \pi^+\pi^+\pi^-$.

\begin{figure*}[htbp]
  \centering
\subfigure[]{\includegraphics[width=0.48\linewidth]{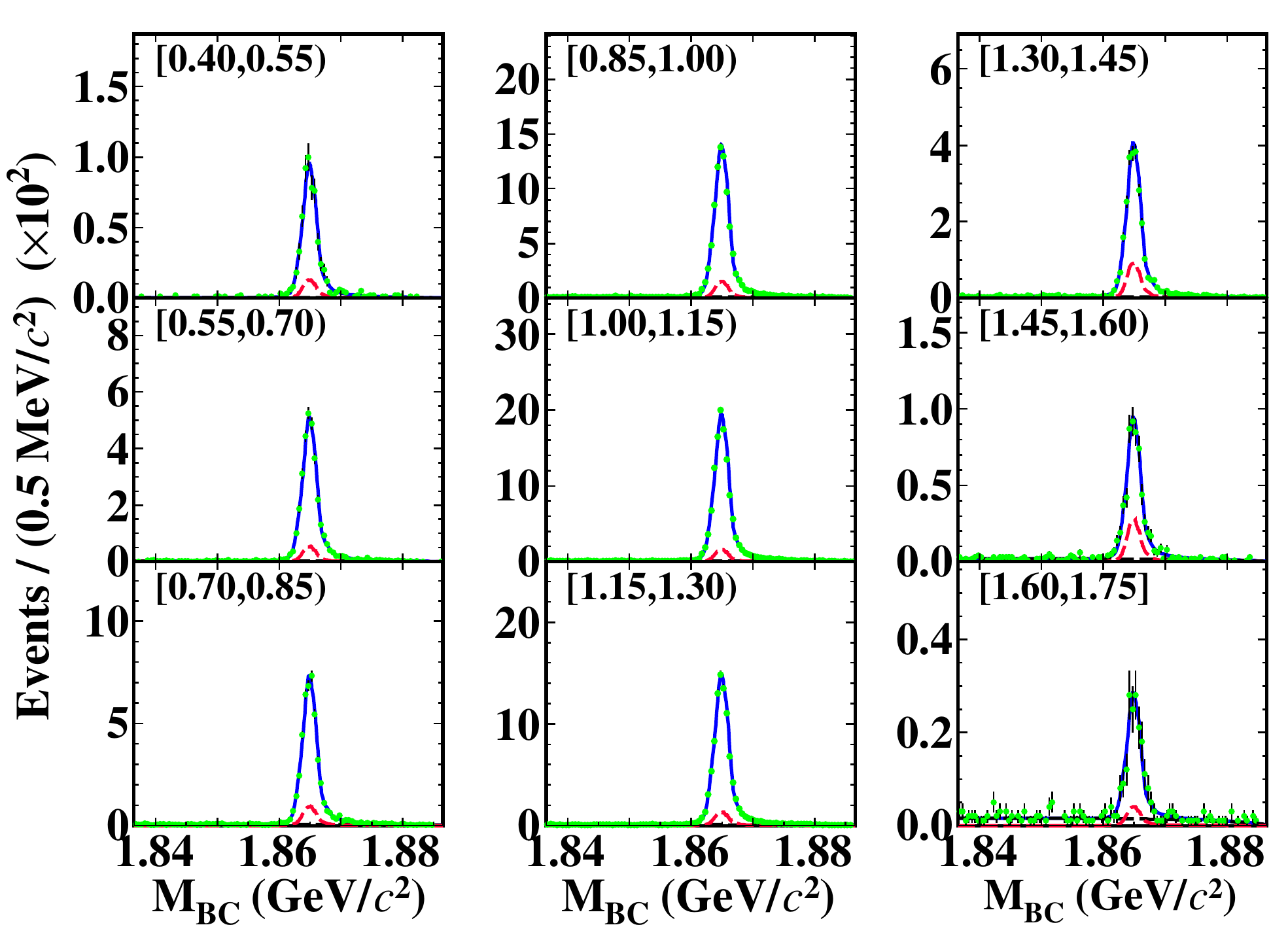}}
\subfigure[]{\includegraphics[width=0.48\linewidth]{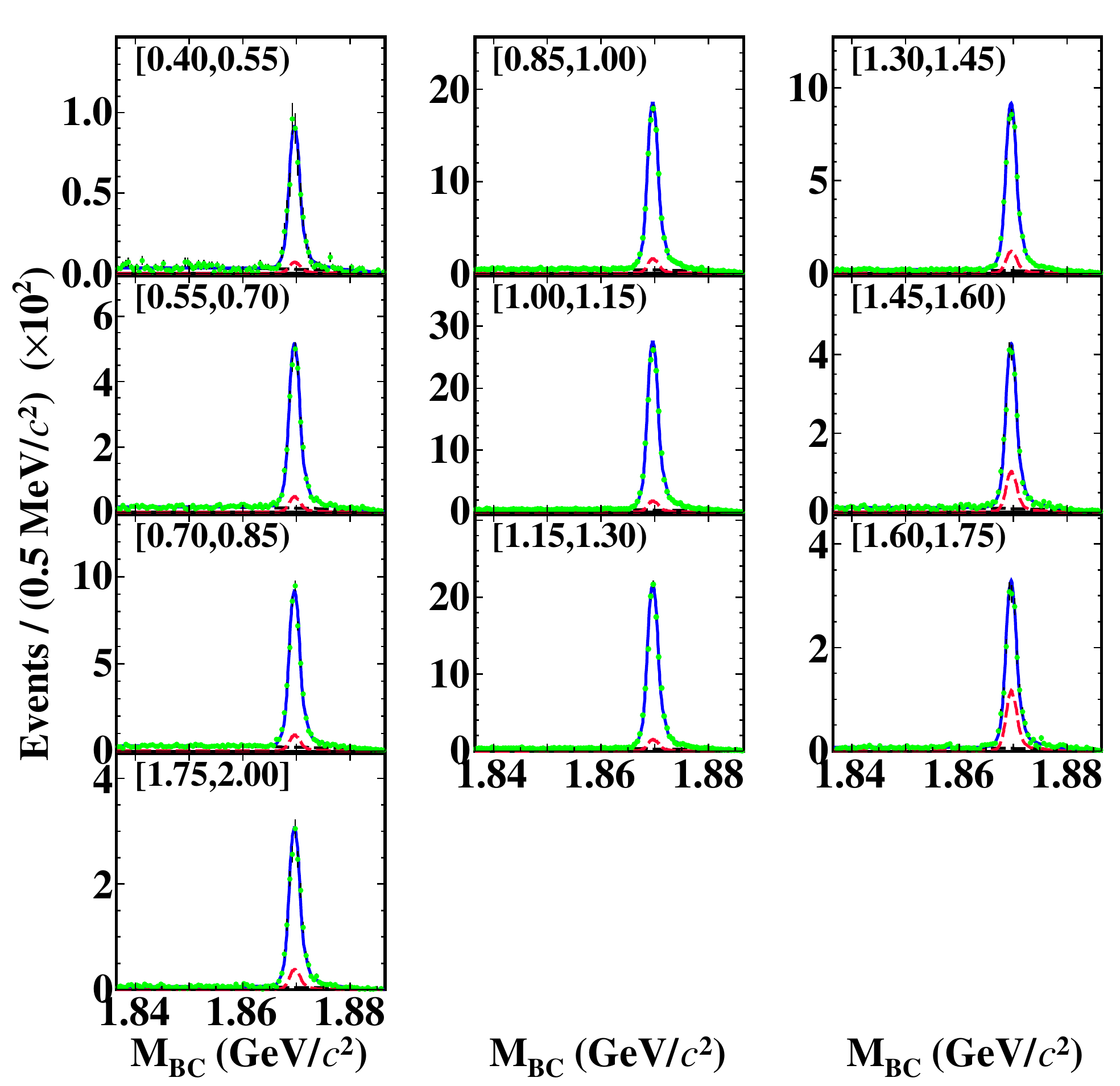}}
  \caption{Fits to the $M_{\rm BC}$ distributions of the tag side when
    the candidates for (a) $D^0\to \pi^+\pi^+\pi^-X$ and (b) $D^+\to
    \pi^+\pi^+\pi^-X$ are found in various reconstructed $M_{3\pi}$
    intervals (in unit of GeV/$c^2$)  indicated on each plot.
The points with error bars are data, the blue solid curves are the fit results,
the pink dashed curves are the peaking background summing over Mis-ID $K$ BKG,
Mis-ID $\mu$ BKG, Mis-ID $e$ BKG, and $K_S^0$ BKG,
and the black dashed curves are the fitted combinatorial backgrounds.}
\label{fig:datafit_MassBC0}
\end{figure*}

The signal yields in each $M_{3\pi}$ bin are determined by fits to the
$M_{\rm BC}$ distributions of the ST side when the candidates for
$D^0\to \pi^+\pi^+\pi^-X$ and $D^+\to \pi^+\pi^+\pi^-X$ are found.
Figures \ref{fig:datafit_MassBC0}(a) and \ref{fig:datafit_MassBC0}(b)
show the $M_{\rm BC}$ distributions of the accepted candidates for
$D^0\to \pi^+\pi^+\pi^-X$ and $D^+\to \pi^+\pi^+\pi^-X$.  The fits to
these $M_{\rm BC}$ distributions are similar to those of the ST
side. Because the ``wrong tag'' background events do not form peaking
background in the $M_{\rm BC}$ distribution of the tag side, an ARGUS
function is used to describe the ``wrong tag'' background.  In these
fits, however, the parameters of the Gaussian functions and the ARGUS
functions are fixed to the values from the $M_{\rm BC}$ fits of the ST
side.  In addition, since the Mis-ID $K$ BKG, Mis-ID $\mu$ BKG, Mis-ID
$e$ BKG, and $K_S^0$ BKG components can peak in the distribution of
$M_{\rm BC}$, the contributions of those backgrounds must be
subtracted. The yields and shapes of the Mis-ID $K$ BKG, Mis-ID $\mu$
BKG, Mis-ID $e$ BKG, and $K_S^0$ BKG components are fixed based on the
fits to the $M_{\rm BC}$ distributions of the background events from
the inclusive MC sample, after taking into account the differences of
the rates of mis-identifying various particles as charged pions
between data and MC simulation.  The fixed background yields and the
signal yields in various $M_{3\pi}$ intervals for $D^0\to
\pi^+\pi^+\pi^-X$ and $D^+\to \pi^+\pi^+\pi^-X$ are summarized in
Tables \ref{DT1} and \ref{DT2}, respectively.

\begin{table*}[htbp]
\begin{center}
\caption{Signal yields ($N_{\rm obs}$) of $D^0 \to \pi^{+}\pi^{+}\pi^{-}X$ observed from data in various reconstructed $M_{3\pi}$ intervals.  The numbers of background events are estimated by the inclusive MC sample, whose integrated luminosity is four times that of data.  The uncertainties of $N_{\rm{Mis-ID}\ \emph{K}\ \rm{BKG}}$, $N_{\rm{Mis-ID}\ \mu\ \rm{BKG}}$, $N_{\rm{Mis-ID}\ e\ \rm{BKG}}$, and $N_{K_S^0\ \rm{BKG}}$ include statistical uncertainties and the uncertainties of individual differences of the particle mis-identification rates between data and MC simulation. The uncertainties of $N_{\rm obs}$ are statistical only. \label{DT1}}
\begin{tabular}{c|r@{\,}c@{\,}l r@{\,}c@{\,}l r@{\,}c@{\,}l r@{\,}c@{\,}l r@{\,}c@{\,}l}
\hline
\hline
$M_{3\pi}$ (GeV/$c^2$)  &\multicolumn{3}{c}{$N_{\rm{Mis-ID}\ \emph{K}\ \rm{BKG}}$}  &\multicolumn{3}{c}{$N_{\rm{Mis-ID}\ \mu\ \rm{BKG}}$ } &\multicolumn{3}{c}{$N_{\rm{Mis-ID}\ e\ \rm{BKG}}$} &\multicolumn{3}{c}{$N_{K_S^0\ \rm{BKG}}$} &\multicolumn{3}{c}{$N_{\rm obs}$}\\
\hline
$[0.40,0.55)$ &$29.1$&$\pm$&$6.9$   &$0.0$&$\pm$&$0.0$    &$4.7$&$\pm$&$0.9$    &$45.5$&$\pm$&$3.5$   &$524.4$&$\pm$&$24.8$      \\
$[0.55,0.70)$ &$78.1$&$\pm$&$15.8$  &$3.1$&$\pm$&$1.0$    &$26.7$&$\pm$&$2.3$   &$221.0$&$\pm$&$7.8$  &$2878.7$&$\pm$&$57.3$     \\
$[0.70,0.85)$ &$81.5$&$\pm$&$14.7$  &$21.2$&$\pm$&$2.6$   &$32.8$&$\pm$&$2.8$   &$428.1$&$\pm$&$10.8$ &$4017.2$&$\pm$&$69.6$     \\
$[0.85,1.00)$ &$85.9$&$\pm$&$14.3$  &$73.8$&$\pm$&$4.8$   &$43.2$&$\pm$&$3.4$   &$721.4$&$\pm$&$14.1$ &$7824.8$&$\pm$&$96.0$     \\
$[1.00,1.15)$ &$58.8$&$\pm$&$9.4$   &$127.6$&$\pm$&$6.3$  &$35.3$&$\pm$&$3.2$   &$766.6$&$\pm$&$14.5$ &$11221.3$&$\pm$&$111.9$   \\
$[1.15,1.30)$ &$13.8$&$\pm$&$2.9$   &$99.3$&$\pm$&$5.4$   &$21.3$&$\pm$&$2.5$   &$670.2$&$\pm$&$13.5$ &$8540.3$&$\pm$&$98.8$     \\
$[1.30,1.45)$ &$15.0$&$\pm$&$2.9$   &$49.4$&$\pm$&$3.8$   &$11.2$&$\pm$&$1.9$   &$485.0$&$\pm$&$11.5$ &$1966.5$&$\pm$&$51.8$     \\
$[1.45,1.60)$ &$6.4$&$\pm$&$1.6$    &$17.6$&$\pm$&$2.2$   &$6.2$&$\pm$&$1.4$    &$135.3$&$\pm$&$6.1$  &$417.7$&$\pm$&$25.3$      \\
$[1.60,1.75]$ &$2.8$&$\pm$&$1.0$    &$0.6$&$\pm$&$0.4$    &$0.0$&$\pm$&$0.0$    &$21.1$&$\pm$&$2.4$   &$145.9$&$\pm$&$14.3$      \\
\hline
\hline
\end{tabular}
\end{center}
\end{table*}

\begin{table*}[htp]
\centering
\begin{center}
\caption{Signal yields ($N_{\rm obs}$) of $D^+ \to \pi^{+}\pi^{+}\pi^{-}X$ observed from data in various reconstructed $M_{3\pi}$ intervals.
The numbers of background events are estimated by the inclusive MC sample, whose integrated luminosity is four times that of data. The uncertainties of $N_{\rm{Mis-ID}\ \emph{K}\ \rm{BKG}}$, $N_{\rm{Mis-ID}\ \mu\ \rm{BKG}}$, $N_{\rm{Mis-ID}\ e\ \rm{BKG}}$, and $N_{K_S^0\ \rm{BKG}}$ include statistical uncertainties and the uncertainties of individual differences of the particle mis-identification rates between data and MC simulation. The uncertainties of $N_{\rm obs}$ are statistical only.\label{DT2}}
\begin{tabular}{c|r@{\,}c@{\,}l r@{\,}c@{\,}l r@{\,}c@{\,}l r@{\,}c@{\,}l r@{\,}c@{\,}l}
\hline
\hline
$M_{3\pi}$ (GeV/$c^2$)  &\multicolumn{3}{c}{$N_{\rm{Mis-ID}\ \emph{K}\ \rm{BKG}}$}  &\multicolumn{3}{c}{$N_{\rm{Mis-ID}\ \mu\ \rm{BKG}}$ } &\multicolumn{3}{c}{$N_{\rm{Mis-ID}\ e\ \rm{BKG}}$} &\multicolumn{3}{c}{$N_{K_S^0\ \rm{BKG}}$} &\multicolumn{3}{c}{$N_{\rm obs}$}\\
\hline
$[0.40,0.55)$ &$21.6$&$\pm$&$4.5$   &$4.5$&$\pm$&$0.9$    &$6.0$&$\pm$&$0.9$    &$9.7$&$\pm$&$1.2$    &$483.3$&$\pm$&$24.5$       \\
$[0.55,0.70)$ &$126.1$&$\pm$&$20.6$ &$52.4$&$\pm$&$3.2$   &$29.6$&$\pm$&$1.9$   &$76.2$&$\pm$&$3.5$   &$2703.8$&$\pm$&$58.4$      \\
$[0.70,0.85)$ &$254.3$&$\pm$&$37.4$ &$119.6$&$\pm$&$4.9$  &$42.4$&$\pm$&$2.5$   &$127.7$&$\pm$&$4.5$  &$4766.7$&$\pm$&$77.6$      \\
$[0.85,1.00)$ &$506.6$&$\pm$&$66.7$ &$244.4$&$\pm$&$7.1$  &$57.4$&$\pm$&$3.1$   &$151.6$&$\pm$&$4.9$  &$9788.2$&$\pm$&$109.6$      \\
$[1.00,1.15)$ &$535.4$&$\pm$&$61.0$ &$346.7$&$\pm$&$8.2$  &$75.2$&$\pm$&$3.6$   &$131.3$&$\pm$&$4.6$  &$14979.8$&$\pm$&$132.3$    \\
$[1.15,1.30)$ &$365.1$&$\pm$&$35.8$ &$365.4$&$\pm$&$8.2$  &$36.1$&$\pm$&$2.6$   &$123.6$&$\pm$&$4.5$  &$11718.5$&$\pm$&$117.0$    \\
$[1.30,1.45)$ &$292.4$&$\pm$&$25.8$ &$296.1$&$\pm$&$7.2$  &$16.0$&$\pm$&$1.7$   &$107.3$&$\pm$&$4.2$  &$4636.0$&$\pm$&$76.8$      \\
$[1.45,1.60)$ &$320.4$&$\pm$&$24.6$ &$195.2$&$\pm$&$5.8$  &$6.6$&$\pm$&$1.1$    &$91.0$&$\pm$&$3.8$   &$1864.6$&$\pm$&$52.8$      \\
$[1.60,1.75)$ &$557.3$&$\pm$&$34.8$ &$84.4$&$\pm$&$3.7$   &$3.0$&$\pm$&$0.7$    &$47.5$&$\pm$&$2.8$   &$1228.0$&$\pm$&$46.0$      \\
$[1.75,2.00]$&$157.4$&$\pm$&$12.1$ &$15.5$&$\pm$&$1.6$   &$0.6$&$\pm$&$0.3$    &$57.0$&$\pm$&$3.0$   &$1559.3$&$\pm$&$44.1$      \\
\hline
\hline
\end{tabular}
\end{center}
\end{table*}

The efficiency matrix $\epsilon_{ij} = N_{\rm reco}^{ij}/N_{\rm prod}^j$ is determined based on signal MC events of all known exclusive $D^{0(+)}$ decays that contain a $\pi^+ \pi^+ \pi^-$ combination,
where $N_{\rm reco}^{ij}$ is the number of signal MC events generated in the $j$-th $M_{3\pi}$ interval and
reconstructed in the $i$-th interval.
The matrix elements of $\epsilon_{ij}$ for $D^0\to \pi^+\pi^+\pi^-X$ and $D^+\to \pi^+\pi^+\pi^-X$
are summarized in Tables~\ref{eff1} and~\ref{eff2}, respectively.

\begin{table}[htbp]
\centering
\setlength{\tabcolsep}{0mm}
\caption{ Efficiency matrix $\epsilon_{ij}$ (in percent) for $D^0 \to \pi^{+}\pi^{+}\pi^{-}X$, where $i$ denotes the reconstructed $M_{3\pi}$ interval
and $j$ denotes the produced $M_{3\pi}$ interval.
The relative statistical uncertainties of the diagonal efficiencies of the matrix are no more than 0.7\%.
}
\begin{center}
\begin{tabular}{c|ccccccccc}
\hline
\hline
$\epsilon_{ij}$                & 1  & 2 & 3 & 4 & 5 & 6 & 7 & 8 & 9 \\\hline
1	&27.74	&0.79	&0.08	&0.03	&0.00	&0.01	&0.03	&0.00	&0.00	\\
2	&2.82	&27.99	&1.01	&0.24	&0.06	&0.02	&0.03	&0.08	&0.00	\\
3	&0.70	&0.95	&25.70	&0.90	&0.16	&0.05	&0.12	&0.34	&0.27	\\
4	&0.69	&0.62	&0.96	&32.65	&1.27	&0.25	&0.20	&0.65	&0.54	\\
5	&0.44	&0.45	&0.33	&1.25	&42.40	&2.07	&0.78	&0.98	&0.54	\\
6	&0.13	&0.09	&0.11	&0.20	&1.07	&47.82	&3.21	&0.81	&0.00	\\
7	&0.00	&0.05	&0.03	&0.06	&0.14	&0.90	&40.78	&1.74	&0.54	\\
8	&0.00	&0.00	&0.00	&0.01	&0.01	&0.05	&0.28	&22.28	&0.80	\\
9	&0.00	&0.00	&0.00	&0.00	&0.00	&0.01	&0.00	&0.34	&20.57	\\
\hline
\hline
\end{tabular}
\label{eff1}
\end{center}
\end{table}

\begin{table}[htbp]
\centering
\setlength{\tabcolsep}{0mm}
\begin{center}
\caption{ Efficiency matrix $\epsilon_{ij}$ (in percent) for $D^+ \to \pi^{+}\pi^{+}\pi^{-}X$, where $i$ denotes the reconstructed $M_{3\pi}$ interval
and $j$ denotes the produced $M_{3\pi}$ interval. The relative statistical uncertainties of the diagonal efficiencies of the matrix are no more than 0.5\%.}
\begin{tabular}{c|cccccccccc}
\hline
\hline
$\epsilon_{ij}$                & 1  & 2 & 3 & 4 & 5 & 6 & 7 & 8 & 9 & 10 \\\hline
1   &22.13  &0.48   &0.11   &0.04   &0.03   &0.03   &0.02   &0.01   &0.00   &0.00   \\
2   &2.18   &24.02  &0.99   &0.25   &0.18   &0.13   &0.11   &0.08   &0.01   &0.00   \\
3   &0.99   &1.71   &22.44  &1.30   &0.41   &0.25   &0.29   &0.10   &0.00   &0.01   \\
4   &0.90   &1.24   &1.90   &30.31  &2.14   &0.83   &0.66   &0.13   &0.04   &0.01   \\
5   &0.81   &1.01   &1.49   &2.53   &39.52  &3.35   &1.33   &0.27   &0.03   &0.01   \\
6   &0.47   &0.71   &1.03   &1.65   &3.39   &47.30  &3.87   &0.35   &0.04   &0.00   \\
7   &0.46   &0.50   &0.75   &1.27   &1.99   &2.33   &50.08  &2.14   &0.34   &0.03   \\
8   &0.15   &0.22   &0.23   &0.36   &0.38   &0.18   &0.76   &54.16  &2.29   &0.19   \\
9   &0.00   &0.00   &0.00   &0.00   &0.01   &0.01   &0.03   &1.28   &56.79  &1.18   \\
10  &0.02   &0.00   &0.00   &0.00   &0.00   &0.00   &0.01   &0.05   &0.71   &55.00  \\
\hline
\hline
\end{tabular}
\label{eff2}
\end{center}
\end{table}

\subsection{Branching fractions}

With the obtained yields of the inclusive decays $D^{0}(D^{+})\to \pi^+ \pi^+ \pi^- X$ produced in the $i$-th $M_{3\pi}$ interval,
the partial decay branching fraction is determined by Eq.~(\ref{branch}).

The measurement of the branching fraction of $D^0\to \pi^+\pi^+\pi^-X$
is affected by the QC effect in neutral $D$ decays, as discussed in
Sec.~\ref{sec:method}.  In this work, the QC factors in the various
$M_{3\pi}$ intervals are estimated by using the $CP$-even ($+$) tag of
$D^0\to K^+K^-$ and $CP$-odd ($-$) tag of $D^0\to K^0_S\pi^0$ with a
similar analysis procedure as the one in the data analysis
procedure presented above.  The quantities used and the
results are summarized in Table~\ref{FQC}.

\begin{table*}[htp]
\centering
\begin{center}
\caption{Quantities of $S_{\rm measured}^{\pm}$, $M_{\rm measured}^{\pm}$, $C_f$, $f_{CP+}$ and ${f}_{\rm QC}^{\rm corr}$ for $D^0 \to \pi^{+}\pi^{+}\pi^{-}X$ in various reconstructed $M_{3\pi}$ intervals.}
\begin{tabular}{c|cccr@{\,}c@{\,}l r@{\,}c@{\,}l r@{\,}c@{\,}l r@{\,}c@{\,}l}
\hline
\hline
$i$  &$S_{\rm measured}^{-}$ &$S_{\rm measured}^{+}$ &$C_{f}$ (\%)
&\multicolumn{3}{c}{$M_{\rm measured}^{-}$}
&\multicolumn{3}{c}{$M_{\rm measured}^{+}$}
&\multicolumn{3}{c}{$f_{CP+}$}
&\multicolumn{3}{c}{${f}_{\rm QC}^{\rm corr}$}\\
\hline
$1$   &\multirow{9}{*}{$70999\pm$304} &\multirow{9}{*}{59238$\pm$442} &\multirow{9}{*}{$-11.3^{+0.4}_{-0.9}$}
&$ 171.8$&$\pm$&$41.6 $ &$ 161.9$&$\pm$&$46.0 $ &$0.46$&$\pm$&$0.14$    &$1.01$&$\pm$&$0.03$    \\
$2$ & & &   &$1163.4$&$\pm$&$90.3 $ &$ 857.1$&$\pm$&$84.6 $ &$0.53$&$\pm$&$0.05$    &$0.99$&$\pm$&$0.01$    \\
$3$ & & &   &$1729.7$&$\pm$&$224.1$ &$2124.3$&$\pm$&$481.8$ &$0.40$&$\pm$&$0.08$    &$1.02$&$\pm$&$0.02$    \\
$4$ & & &   &$2697.6$&$\pm$&$149.4$ &$2620.2$&$\pm$&$295.8$ &$0.46$&$\pm$&$0.04$    &$1.01$&$\pm$&$0.01$    \\
$5$ & & &   &$3241.0$&$\pm$&$132.0$ &$2798.1$&$\pm$&$161.7$ &$0.49$&$\pm$&$0.03$    &$1.00$&$\pm$&$0.01$    \\
$6$ & & &   &$1776.2$&$\pm$&$79.6 $ &$2108.6$&$\pm$&$89.8 $ &$0.41$&$\pm$&$0.02$    &$1.02$&$\pm$&$0.01$    \\
$7$ & & &   &$ 360.7$&$\pm$&$52.2 $ &$ 547.9$&$\pm$&$72.6 $ &$0.35$&$\pm$&$0.10$    &$1.04$&$\pm$&$0.03$    \\
$8$ & & &   &$ 250.2$&$\pm$&$50.5 $ &$ 186.0$&$\pm$&$45.6 $ &$0.52$&$\pm$&$0.18$    &$0.99$&$\pm$&$0.04$    \\
$9$ & & &   &$ 121.8$&$\pm$&$33.8 $ &$  21.5$&$\pm$&$8.1  $   &$0.82$&$\pm$&$0.60$  &$0.93$&$\pm$&$0.12$    \\
\hline
\hline
\end{tabular}
\label{FQC}
\end{center}
\end{table*}

The produced signal yields and the obtained partial branching fractions of $D^0 \to \pi^{+}\pi^{+}\pi^{-}X$ and $D^+ \to \pi^{+}\pi^{+}\pi^{-}X$ in different $M_{3\pi}$ intervals
are presented in Tables~\ref{BF1} and~\ref{BF2}, respectively.

\begin{table}[htbp]
\centering
\begin{center}
\caption{The produced signal yields and the obtained partial branching fractions of $D^0 \to \pi^{+}\pi^{+}\pi^{-}X$ in different reconstructed $M_{3\pi}$ intervals. ${d\mathcal B}_{\rm sig}^{CP}$ is the partial decay branching fraction corrected by the QC factor, where ${d\mathcal B}_{\rm sig}^{\rm corr}=f^{\rm corr}_{\rm QC}\times {d\mathcal B}_{\rm sig}$ according to Eq.~(\ref{equ:QCcorrect}).}
\label{BF1}
\begin{tabular}{c|r@{\,}c@{\,}l r@{\,}c@{\,}l r@{\,}c@{\,}l}
\hline
\hline
$i$   &\multicolumn{3}{c}{$N_{\rm prod}$} &\multicolumn{3}{c}{${d\mathcal B}_{\rm sig}$} &\multicolumn{3}{c}{${d\mathcal B}_{\rm sig}^{\rm corr}$ (\%)}\\
\hline
$1$  &$1541.3$&$\pm$&$89.9$     &0.28$$&$\pm$&$$0.02    &0.28$$&$\pm$&$$0.02  \\
$2$  &$9349.1$&$\pm$&$206.0$    &1.71$$&$\pm$&$$0.04    &1.70$$&$\pm$&$$0.04  \\
$3$  &$14235.8$&$\pm$&$271.8$   &2.60$$&$\pm$&$$0.05    &2.66$$&$\pm$&$$0.05  \\
$4$  &$22130.5$&$\pm$&$295.0$   &4.04$$&$\pm$&$$0.05    &4.08$$&$\pm$&$$0.05  \\
$5$  &$24638.2$&$\pm$&$264.9$   &4.50$$&$\pm$&$$0.05    &4.51$$&$\pm$&$$0.05  \\
$6$  &$16850.4$&$\pm$&$207.4$   &3.07$$&$\pm$&$$0.04    &3.14$$&$\pm$&$$0.04  \\
$7$  &$4228.6$&$\pm$&$127.5$    &0.77$$&$\pm$&$$0.02    &0.80$$&$\pm$&$$0.02  \\
$8$  &$1730.9$&$\pm$&$113.7$    &0.32$$&$\pm$&$$0.02    &0.31$$&$\pm$&$$0.02  \\
$9$  &$676.1$&$\pm$&$69.6$      &0.12$$&$\pm$&$$0.01    &0.11$$&$\pm$&$$0.01  \\
\hline
Total &$95381.0$&$\pm$&$598.9$	&&$-$&   &$17.60$&$\pm$&$0.11$\\
\hline
\hline
\end{tabular}
\end{center}
\end{table}

\begin{table}[htbp]
\centering
\begin{center}
\caption{The produced signal yields and the obtained partial branching fractions of $D^+ \to \pi^{+}\pi^{+}\pi^{-}X$ in different reconstructed $M_{3\pi}$ intervals. \label{BF2}
}
\begin{tabular}{c|r@{\,}c@{\,}l r@{\,}c@{\,}l}
\hline
\hline
$i$  &\multicolumn{3}{c}{$N_{\rm prod}$}  &\multicolumn{3}{c}{${d\mathcal B}_{\rm sig}$ (\%)}\\
\hline
$1$  &$ 1747.1$&$\pm$&$111.1$   &0.22$$&$\pm$&$$0.01  \\
$2$  &$ 9683.3$&$\pm$&$245.1$   &1.19$$&$\pm$&$$0.03  \\
$3$  &$17890.3$&$\pm$&$349.6$   &2.20$$&$\pm$&$$0.04  \\
$4$  &$27671.6$&$\pm$&$366.3$   &3.41$$&$\pm$&$$0.05  \\
$5$  &$33224.6$&$\pm$&$340.2$   &4.09$$&$\pm$&$$0.04  \\
$6$  &$20383.9$&$\pm$&$251.5$   &2.51$$&$\pm$&$$0.03  \\
$7$  &$ 5772.7$&$\pm$&$155.4$   &0.71$$&$\pm$&$$0.02  \\
$8$  &$ 2661.8$&$\pm$&$97.8$    &0.33$$&$\pm$&$$0.01  \\
$9$  &$ 2032.0$&$\pm$&$81.1$    &0.25$$&$\pm$&$$0.01  \\
$10$ &$ 2803.0$&$\pm$&$80.2$    &0.35$$&$\pm$&$$0.01  \\
\hline
Total &$123870.2$&$\pm$&$744.7$  &$15.25$&$\pm$&$0.09$\\
\hline
\hline
\end{tabular}
\end{center}
\end{table}

\section{Systematic uncertainties}
\label{SYS}
Benefiting from the DT method, the branching fraction measurements are
insensitive to the selection criteria of the ST $\bar D$ candidates.
The systematic uncertainties in the measurements of the branching
fractions of $D^0\to \pi^+\pi^+\pi^-X$ and $D^+\to \pi^+\pi^+\pi^-X$
are
discussed below.

Both the ST and DT yields are determined from the fits to the
individual $M_{\rm BC}$ distributions.  The fits to the DT candidates
are performed by using the same fit strategy as the fits to the ST
candidates, with the parameters of the smeared Gaussian function and
the ARGUS background function derived from the corresponding fits to
the ST candidates.  In this case, the fitted DT yields are correlated
to those of the fitted ST yields.  Therefore, the systematic
uncertainties in the yields of the ST $\bar D$ mesons are canceled in
the branching fraction measurements.

The tracking and PID efficiencies of $\pi^\pm$ are studied with the DT hadronic $D\bar D$ events.
The averaged data/MC differences of $\pi^\pm$ tracking and PID efficiencies, weighted by the corresponding momentum spectra of signal MC events, are $0.62\%$ and $0.17\%$, respectively.
After correcting the MC efficiencies to data by these averaged data/MC differences, the systematic uncertainties of tracking and PID efficiencies for the three charged pions are assigned as 0.80\% and 0.50\% for  $D^0\to \pi^+\pi^+\pi^-X$ and $D^+\to \pi^+\pi^+\pi^-X$, respectively.

The detection efficiencies of $D^{0(+)}\to \pi^+\pi^+\pi^-X$ are obtained from the signal MC sample including all known decays with three charged pions. The relevant systematic uncertainties are estimated by varying the input branching fractions of exclusive decays within $\pm 1\sigma$. The maximum changes of the detection efficiencies, 0.13\% and 0.56\%, are assigned as the corresponding systematic uncertainties for $D^0\to \pi^+\pi^+\pi^-X$ and $D^+\to \pi^+\pi^+\pi^-X$, respectively.

The uncertainties due to the limited signal DT MC samples are calculated by~\cite{MC}
\begin{equation}\label{MCsys}
C_{ij}^{\rm sys}=(\frac{1}{N_{\rm ST}})^2\sum_{\alpha\beta}N_{\rm obs}^{\alpha}N_{\rm obs}^{\beta}\rm{Cov}(\epsilon_{\emph{i}\alpha}^{-1},\epsilon_{\emph{j}\beta}^{-1}),
\end{equation}
where the covariances of the inverse efficiency matrix elements are given by
\begin{equation}\label{Cov}
\rm{Cov}(\epsilon_{\alpha\beta}^{-1},\epsilon_{\emph{ab}}^{-1})=\sum_{\emph{ij}}(\epsilon_{\alpha \emph{i}}^{-1}\epsilon_{\emph{ai}}^{-1})[\sigma(\epsilon_{\emph{ij}})]^2(\epsilon_{\emph{j}\beta}^{-1}\epsilon_{\emph{jb}}^{-1}).
\end{equation}
The corresponding systematic uncertainties are assigned to be 0.31\% and 0.23\% for $D^0\to \pi^+\pi^+\pi^-X$ and $D^+\to \pi^+\pi^+\pi^-X$, respectively.

The efficiencies of mis-identifying $e^\pm$, $\mu^\pm$ and $K^\pm$ as $\pi^\pm$ are studied with the
$e^+e^-\to\gamma e^+e^-$ events, the $e^+e^-\to\gamma \mu^+\mu^-$ events, and the DT hadron $D\bar D$ events, respectively. The averaged data/MC differences of the efficiencies of mis-identifying $e^\pm$ ($\mu^\pm$) as $\pi^\pm$, weighted by the two-dimensional (momentum and $\cos\theta$) distributions of signal MC events, are $12\%$ and $5\%$, respectively.
The averaged data/MC differences of the efficiencies of mis-identifying $K^\pm$ as $\pi^\pm$, weighted by the corresponding momentum spectra of signal MC events, are no more than $53\%$. Here, the large data/MC differences in some $M_{3\pi}$ intervals are mainly due to \textcolor{blue}{extremely} small rates of mis-identifying $K^\pm$ as $\pi^\pm$, which are in the range of $(0.05-0.72) \%$ for $p_{K}<0.7$ GeV/$c$.
After correcting these mis-identification efficiencies to data by these averaged data/MC differences,
the associated systematic uncertainties are estimated by varying the fixed background yields within their individual uncertainties.
Their effects on the measured branching fractions are 0.13\% and 0.18\% for  $D^0\to \pi^+\pi^+\pi^-X$ and $D^+\to \pi^+\pi^+\pi^-X$, respectively.

The yields of background with correct $\bar D$ tag but wrong $D$ signal are estimated using the inclusive MC sample.
The relevant systematic uncertainty is evaluated by varying the world average branching fractions of the top five background components (which account for more than 52\% of all background contributions) of
the decays in $D^{0(+)}\to \pi^+\pi^+\pi^-X$ within $\pm 1\sigma$. The changes of the branching fractions are assigned as the corresponding systematic uncertainties, which are 0.09\% and 0.20\% for $D^0\to \pi^+\pi^+\pi^-X$ and $D^+\to \pi^+\pi^+\pi^-X$, respectively. The effects of the remaining background components are also examined similarly and found to be negligible.

The $K_S^0$ veto is considered in three aspects.  The first systematic
uncertainty comes from the requirements of rejecting ($K_S^0$ BKG1).
The associated systematic uncertainty is estimated by altering the
nominal veto window of $|M_{\pi^+\pi^-}-0.4977|>0.030~$~GeV/$c^2$ by
$\pm 5$ MeV/$c^2$, which corresponds to about $1\sigma$ of the fitted $K^0_S$ mass resolution. The largest change of the re-measured branching
fraction is taken as the systematic uncertainties, which are 0.23\%
and 0.06\% for $D^0\to \pi^+\pi^+\pi^-X$ and $D^+\to
\pi^+\pi^+\pi^-X$, respectively.  The second systematic uncertainty
comes from the requirements of rejecting ($K_S^0$ BKG2), and it is
estimated by varying the background yields by
$\pm$1.5\% in the $M_{\rm BC}$ fit.
Here, the 1.5\% corresponds to the data/MC difference of the $K^0_S$ reconstruction efficiencies between data and MC simulation, estimated with the control sample of $J/\psi \to K^*(892)^{\mp}K^{\pm}$ and $J/\psi \to \phi K_S^0K^{\pm}\pi^{\mp}$~\cite{bes3-klev}.
The changes in the re-measured
branching fractions are assigned as the systematic uncertainties,
which are 0.11\% and 0.17\% for $D^0\to \pi^+\pi^+\pi^-X$ and $D^+\to
\pi^+\pi^+\pi^-X$, respectively.  The third systematic uncertainty
comes from the requirements of rejecting ($K_S^0$ BKG3), and it is
estimated by altering the nominal veto window of
$|M_{\pi^+\pi^-}-0.4977|>0.080~$~GeV/$c^2$ by $\pm$5~MeV/$c^2$. The
largest change of the re-measured branching fraction is taken as the
systematic uncertainty, which is 0.28\% for $D^0\to \pi^+\pi^+\pi^-X$.
Adding these three items in quadrature yields the systematic
uncertainties due to $K_S^0$ veto to be 0.38\% and 0.18\% for $D^0\to
\pi^+\pi^+\pi^-X$ and $D^+\to \pi^+\pi^+\pi^-X$, respectively.

The systematic uncertainties due to $M_{3\pi}$ divisions are estimated by increasing or decreasing
the interval number by 50\%. The larger differences of the re-measured branching fractions to the nominal results
are assigned as the systematic uncertainties, which are 0.34\% and 0.20\% for $D^0\to \pi^+\pi^+\pi^-X$ and $D^+\to \pi^+\pi^+\pi^-X$, respectively.

The systematic uncertainty due to the correction factor of QC effect in $D^0\to \pi^+ \pi^+ \pi^-X$ decays is determined by the residual uncertainty of $f_{\rm QC}^{\rm corr}$, which are summarized in Table~\ref{FQC}. After weighting by the corresponding numbers of the inclusive decays $D^{0}\to \pi^+ \pi^+ \pi^- X$ produced in each $i$-th $M_{3\pi}$ interval, 0.42\% is taken as the systematic uncertainty.

Assuming all systematic uncertainties to be uncorrelated and adding them in quadrature,
we obtain the total systematic uncertainties in the measurements of the branching fractions of $D^0\to \pi^+\pi^+\pi^-X$ and $D^+\to \pi^+\pi^+\pi^-X$ to be 1.23\% and 1.18\%, respectively.
Table~\ref{tab:sys} summarizes the systematic uncertainties discussed above.

\begin{table*}[htbp]
\centering
\caption{Relative systematic uncertainties (in percent) in the measurements of the branching fractions
of $D^0\to \pi^+\pi^+\pi^-X$ and $D^+\to \pi^+\pi^+\pi^-X$.
}
\begin{tabular}{lcc}
  \hline
  \hline
  Source  &$D^0\to \pi^+\pi^+\pi^-X$ &$D^+\to \pi^+\pi^+\pi^-X$  \\
  \hline
  $\pi^\pm$ tracking                 &0.80       &0.80      \\
  $\pi^\pm$ PID                      &0.50       &0.50     \\
  Efficiency estimate                &0.13       &0.56      \\
  MC statistics                      &0.31       &0.23      \\
  Mis-identification efficiencies    &0.13       &0.18      \\
  Background estimate                &0.09       &0.20      \\
  $K_S^0$ vetoes                     &0.38       &0.18      \\
  QC correction factor               &0.42       &-        \\
  Binning scheme                     &0.34       &0.20      \\
  \hline
  Total                              &1.23       &1.18      \\
  \hline
  \hline
\end{tabular}
\label{tab:sys}
\end{table*}

\section{Summary}

By analyzing 2.93~fb$^{-1}$ of $e^+e^-$ collision data taken at $\sqrt{s}$ = 3.773~GeV, we have measured the
branching fractions of the inclusive decays  $D^0\to\pi^+\pi^+\pi^-X$ and $D^+\to\pi^+\pi^+\pi^-X$ for the first time.
The results are
\begin{equation}
  {\mathcal B}(D^0\to \pi^+\pi^+\pi^- X)=  (17.60\pm 0.11\pm 0.22)\%,\nonumber
\end{equation}
and
\begin{equation}
  {\mathcal B}(D^+\to \pi^+\pi^+\pi^- X)=  (15.25\pm 0.09\pm 0.18)\%,\nonumber
\end{equation}
where the first uncertainties are statistical and the second are
systematic. They are consistent with the sums of the branching
fractions of the known decay modes as summarized in
Appendix~\ref{sec:app} within about $\pm 3\sigma$. The partial
branching fraction in the interval [1.75, 2.00] GeV/$c^2$ is
consistent with the branching fraction of the exclusive hadronic decay
of $D^+\to \pi^+\pi^+\pi^-$, which is $(0.327\pm0.018)\%$ according to
the PDG~\cite{pdg2020}. These results indicate that there is little
room for possible missing $D^{0}(D^+)$ decays containing
$\pi^+\pi^+\pi^-$. The measured total and partial branching fractions
of $D^0(D^+)\to \pi^+\pi^+\pi^-X$ are important inputs for LFU tests
for the semileptonic $B$ decays.

\section{Acknowledgement}

The BESIII collaboration thanks the staff of BEPCII and the IHEP computing center for their strong support. This work is supported in part by National Key R\&D Program of China under Grants Nos. 2020YFA0406400, 2020YFA0406300; National Natural Science Foundation of China (NSFC) under Grants Nos. 11875170, 12035009, 11635010, 11735014, 11835012, 11935015, 11935016, 11935018, 11961141012, 12022510, 12025502, 12035013, 12061131003, 12192260, 12192261, 12192262, 12192263, 12192264, 12192265; the Chinese Academy of Sciences (CAS) Large-Scale Scientific Facility Program; the CAS Center for Excellence in Particle Physics (CCEPP); Joint Large-Scale Scientific Facility Funds of the NSFC and CAS under Grant No. U1832207; CAS Key Research Program of Frontier Sciences under Grants Nos. QYZDJ-SSW-SLH003, QYZDJ-SSW-SLH040; 100 Talents Program of CAS; The Institute of Nuclear and Particle Physics (INPAC) and Shanghai Key Laboratory for Particle Physics and Cosmology; ERC under Grant No. 758462; European Union's Horizon 2020 research and innovation programme under Marie Sklodowska-Curie grant agreement under Grant No. 894790; German Research Foundation DFG under Grants Nos. 443159800, 455635585, Collaborative Research Center CRC 1044, FOR5327, GRK 2149; Istituto Nazionale di Fisica Nucleare, Italy; Ministry of Development of Turkey under Grant No. DPT2006K-120470; National Science and Technology fund; National Science Research and Innovation Fund (NSRF) via the Program Management Unit for Human Resources \& Institutional Development, Research and Innovation under Grant No. B16F640076; Olle Engkvist Foundation under Grant No. 200-0605; STFC (United Kingdom); Suranaree University of Technology (SUT), Thailand Science Research and Innovation (TSRI), and National Science Research and Innovation Fund (NSRF) under Grant No. 160355; The Royal Society, UK under Grants Nos. DH140054, DH160214; The Swedish Research Council; U. S. Department of Energy under Grant No. DE-FG02-05ER41374.

\appendix
\twocolumngrid
\setcounter{table}{0}
\setcounter{figure}{0}

\section{Branching fractions of the known exclusive $D^0(D^+)$ decays involving $\pi^+\pi^+\pi^-$}
\label{sec:app}
Tables~\ref{signal_D0} and \ref{signal_Dp} show the intermediate and final states that contribute to the inclusive decays $D^0 \to \pi^{+}\pi^{+}\pi^{-}X$
and $D^+ \to \pi^{+}\pi^{+}\pi^{-}X$ as well as the known branching fractions. The known branching fractions of $D^0 \to \pi^{+}\pi^{+}\pi^{-}X$ and $D^+ \to \pi^{+}\pi^{+}\pi^{-}X$ are ${\mathcal B}(D^0\to \pi^+\pi^+\pi^- X)=  (16.05\pm 0.47)\%$ and ${\mathcal B}(D^+\to \pi^+\pi^+\pi^- X)=  (14.74\pm0.53)\%$, respectively.

\begin{table*}[htbp]
\centering
\footnotesize
\caption{ The initial and final states contributing to the inclusive decay $D^0 \to \pi^{+}\pi^{+}\pi^{-}X$, along with the known branching fractions.
The branching fractions of the hadronic $D$ decays containing $\bar K^0$ have been obtained by scaling the known branching fractions of $K^0_S$ by a factor of two. Any $\pi^+$ or $\pi^-$ from $K^0_S$ decays have not been included.
The contributions of some decays containing $\eta$, $\eta^{\prime}$,  $\omega$, and $\phi$ have been excluded to avoid overlaps among various decays.
\label{signal_D0}
}
\begin{center}
\begin{tabular}{c|c|c|c|c|c}
\hline
\hline
Initial state                    &$\mathcal{B}^{\rm Initial} (\%)$  &Final state   &$\mathcal{B}^{\rm Final} (\%)$ &Reference &Note     \\
\hline
$K^{-}\pi^{+}\pi^{+}\pi^{-}$     &$8.220\pm0.140$      &$K^{-}\pi^{+}\pi^{+}\pi^{-}$                &$8.168 \pm0.145$    &~\cite{pdg2020}              &None $\omega$\\

$K^{-}\pi^+\omega$               &$3.392\pm0.096$      &$K^{-}\pi^+\pi^+\pi^-X$                         &$3.088\pm0.087$   &~\cite{wuxiao} &Scaled by ${\mathcal B} (\omega\to \pi^+\pi^-X) \sim 91.0\%$\\

$K^{-}\pi^{+}\pi^{+}\pi^{-}\pi^{0}$	 &$4.300\pm0.400$  &$K^{-}\pi^{+}\pi^{+}\pi^{-}\pi^{0}$	     &$0.845 \pm0.409$    &~\cite{pdg2020} &None $\eta$, $\eta^{\prime}$, and $\omega$\\

$\pi^+\pi^{+}\pi^{-}\pi^{+}$         &$0.755\pm0.020$  &$\pi^+\pi^{+}\pi^{-}\pi^{+}$               &$0.755 \pm0.020$    &~\cite{pdg2020} &...               \\

$K^-\pi^+\eta^{\prime}$              &$0.643\pm0.034$  &$K^-\pi^+\pi^+\pi^-X$          &$0.525\pm0.028 $    &~\cite{pdg2020}&Scaled by ${\mathcal B} (\eta^{\prime}\to \pi^+\pi^-X) \sim 81.7\%$              \\

$K^{-}\pi^{+}\eta$                   &$1.880\pm0.050$  &$K^{-}\pi^{+}\pi^+\pi^-X$     &$0.514 \pm0.013$    &~\cite{pdg2020}&Scaled by ${\mathcal B} (\eta\to \pi^+\pi^-X) \sim 27.4\%$                           \\

$K^{*-}\rho^0\pi^+$             &$0.320\pm0.060$  &$K^{*-}\pi^+\pi^-\pi^+$     &$0.320 \pm0.060$    &~\cite{pdg2020}&Scaled by ${\mathcal B} (\rho^0\to \pi^+\pi^-) \sim 100\%$                     \\

$\pi^+\pi^+\pi^-\pi^-\pi^0\pi^0$     &$0.442\pm0.029$  &$\pi^+\pi^+\pi^-\pi^-\pi^0\pi^0$              &$0.317 \pm0.030$    &~\cite{kaikai}&None $\eta$, $\eta^{\prime}$ and $\omega$                      \\

$\pi^+\pi^{+}\pi^{-}\pi^{-}\pi^{0}$  &$0.420\pm0.050$  &$\pi^+\pi^{+}\pi^{-}\pi^{-}\pi^{0}$        &$0.275 \pm0.053$    &~\cite{pdg2020}&None $\eta$, $\eta^{\prime}$, and $\omega$                      \\

$K^{0}\eta^{\prime}$                 &$1.898\pm0.045$  &$K^{0}\pi^+\pi^-\pi^+\pi^-X$       &$0.226\pm0.005$    &~\cite{pdg2020} &Scaled by ${\mathcal B} (\eta^{\prime} \to \pi^+\pi^-\eta)\times{\mathcal B} (\eta\to \pi^+\pi^-X) \sim 11.9\%$    \\

$\bar K^0\rho^0\pi^+\pi^-$           &$0.220\pm0.099$       &$\bar K^0\pi^+\pi^+\pi^-\pi^-$        &$0.220 \pm0.099$    &~\cite{pdg2020}&Scaled by ${\mathcal B}(\rho^0\to \pi^+\pi^-) \sim 100\%$                     \\

$K^{-}\pi^{+}\pi^{0}\eta$            &$0.449\pm0.027$  &$K^{-}\pi^{+}\pi^0\pi^+\pi^-X$&$0.123 \pm0.007$    &~\cite{pdg2020}&Scaled by ${\mathcal B} (\eta\to \pi^+\pi^-X) \sim 27.4\%$                           \\

$\pi^{+}\pi^{-}\omega$               &$0.133\pm0.020$  &$\pi^+\pi^-\pi^{+}\pi^{-}X$   &$0.121 \pm0.018$    &~\cite{pdg2020}&Scaled by ${\mathcal B} (\omega \to \pi^+\pi^-X) \sim 91.0\%$                           \\

$K^{0}\pi^{+}\pi^{-}\pi^{+}\pi^{-}$  &$<0.120$  &$K^{0}\pi^{+}\pi^{-}\pi^{+}\pi^{-}$        &$0.12$  &~\cite{pdg2020}&...              \\

Others &$0.543\pm0.041$        &$\pi^+\pi^+\pi^-X$           &$0.434\pm0.033$ &~\cite{pdg2020,kaikai}&Dominated by $\pi^+\pi^-\eta X, \pi^+\pi^+\pi^-\eta^{\prime} X, \eta\eta^{\prime}$ and $\omega\eta~\sim 80\%$ \\
\hline
Sum    &...    &...  &$16.05\pm0.47$    &...   &...\\
\hline
\hline
\end{tabular}
\end{center}
\end{table*}

\begin{table*}[htbp]
\centering
\footnotesize
\caption{ The initial and final states contributing to the inclusive decay $D^+ \to \pi^{+}\pi^{+}\pi^{-}X$, along with the known branching fractions.
The branching fractions of the hadronic $D$ decays containing $\bar K^0$ have been obtained by scaling the known branching fractions of $K^0_S$ by a factor of two.  Any $\pi^+$ or $\pi^-$ from $K^0_S$ decays have not been included.
The contributions of some decays containing $\eta$, $\eta^{\prime}$, $\omega$, and $\phi$ have been excluded to avoid overlaps among various decays.
\label{signal_Dp}
}
\begin{center}
\begin{tabular}{c|c|c|c|c|c}
\hline
\hline
Initial state     &$\mathcal{B}^{\rm Initial} (\%)$        &Final state      &$\mathcal{B}^{\rm Final} (\%)$       &Reference          &Note \\
\hline
$\bar K^0\pi^+\pi^+\pi^-$       &$6.200\pm0.127$      &$\bar K^0\pi^+\pi^+\pi^-$ &$6.178\pm0.130$    &\cite{pdg2020}&None $\omega$            \\

$\bar K^0\pi^+\omega$           &$1.414\pm0.071$      &$\bar K^0\pi^+\pi^+\pi^-X$        &$1.287 \pm0.053$ &\cite{pdg2020}&Scaled by ${\mathcal B} (\omega \to \pi^+\pi^-X) \sim 91.0\%$                    \\

$\bar K^0\pi^+\pi^+\pi^-\pi^0$  &$3.056\pm0.102$      &$\bar K^0\pi^+\pi^+\pi^-\pi^0$    &$1.198\pm0.117$ &\cite{lanxing}&None $\eta$, $\eta^{\prime}$, and $\omega$               \\

$\pi^+\pi^+\pi^-\pi^0$          &$1.160\pm0.080$      &$\pi^+\pi^+\pi^-\pi^0$            &$0.954  \pm0.083$    &\cite{pdg2020}&None $\eta$, $\eta^{\prime}$, $\omega$ and $\phi$            \\

$\bar K^0\pi^+\eta$             &$2.620\pm0.071$       &$\bar K^0\pi^+\pi^+\pi^-X$        &$0.718 \pm0.019$         &\cite{pdg2020}&Scaled by ${\mathcal B}(\eta\to \pi^+\pi^-X) \sim 27.4\%$                        \\

$K^-\pi^+\pi^+\pi^+\pi^-$       &$0.570\pm0.050$      &$K^-\pi^+\pi^+\pi^+\pi^-$         &$0.570  \pm0.050  $    &\cite{pdg2020}&...    \\

$\pi^+\eta^{\prime}$            &$0.497\pm0.019$      &$\pi^+\pi^+\pi^-X$                &$0.406\pm0.016$    &\cite{pdg2020}&Scaled by ${\mathcal B}(\eta^{\prime}\to \pi^+\pi^-X) \sim 81.7\%$                    \\

$\pi^+\pi^0\phi$                &$  2.3\pm1.0  $          &$\pi^+\pi^0\pi^+\pi^-X$           &$0.362 \pm0.157$ &\cite{pdg2020}&Scaled by ${\mathcal B} (\phi \to \pi^+\pi^-X) \sim 15.6\%$       \\

$\pi^+\pi^0\omega$              &$ 0.39\pm0.09 $       &$\pi^+\pi^0\pi^+\pi^-X$           &$0.355 \pm0.082$   &\cite{pdg2020}&Scaled by ${\mathcal B} (\omega \to \pi^+\pi^-X) \sim 91.0\%$                          \\

$\pi^+\pi^+\pi^-\pi^0\pi^0$     &$1.07\pm0.41$        &$\pi^+\pi^+\pi^-\pi^0\pi^0$       &$0.322  \pm0.445  $ &\cite{pdg2020}&None $\eta$, $\eta^{\prime}$, $\omega$ and $\phi$             \\

$\bar K^0\pi^+\eta^{\prime}$    &$0.380\pm0.030$      &$\bar K^0\pi^+\pi^+\pi^-X$     &$0.311 \pm0.025$         &\cite{pdg2020}&Scaled by ${\mathcal B}(\eta^{\prime}\to \pi^+\pi^-X) \sim 81.7\%$                   \\

$\pi^+\pi^+\pi^-$               &$0.327\pm0.018$      &$\pi^+\pi^+\pi^-$                 &$0.305 \pm0.018$    &\cite{pdg2020}&None $\omega$            \\

$\pi^+\pi^+\pi^-\pi^0\eta$      &$0.388\pm0.032$&$\pi^+\pi^+\pi^-\pi^0\eta$              &$0.251 \pm0.040 $ &\cite{kaikai}&None $\eta$ and $\eta^{\prime}$          \\

$\pi^+\pi^+\pi^-\pi^0\pi^0\pi^0$&$0.347\pm0.031$      &$\pi^+\pi^+\pi^-\pi^0\pi^0\pi^0$  &$0.250 \pm0.032 $ &\cite{kaikai}&None $\eta$            \\

$\pi^+\pi^+\pi^+\pi^-\pi^-$     &$0.166\pm0.016$      &$\pi^+\pi^+\pi^+\pi^-\pi^-$       &$0.166 \pm0.016 $    &\cite{pdg2020}&...            \\

$\pi^+\eta\eta$                 &$0.296\pm0.026$      &$\pi^+\eta\pi^+\pi^-X$            &$0.162 \pm0.014$   &\cite{pdg2020}&Scaled by ${\mathcal B}(\eta\to \pi^+\pi^-X)\times2 \sim 54.8\%$                    \\

$\pi^+\pi^+\pi^+\pi^-\pi^-\pi^0$&$0.238\pm0.022$      &$\pi^+\pi^+\pi^+\pi^-\pi^-\pi^0$  &$0.160 \pm0.023 $ &\cite{kaikai}&None $\eta$          \\

$\pi^+\pi^0\eta^{\prime}$       &$0.16\pm0.05$         &$\pi^+\pi^0\pi^+\pi^-X$          &$0.131\pm0.041$     &\cite{pdg2020}&Scaled by ${\mathcal B}(\eta^{\prime}\to \pi^+\pi^-X) \sim 81.7\%$                    \\

$\pi^+\pi^+\pi^-\eta$           &$0.341\pm0.021$       &$\pi^+\pi^+\pi^-\eta$            &$0.125 \pm0.023$  &\cite{pdg2020}&None $\eta^{\prime}$      \\

$\pi^+\eta$                     &$0.377\pm0.009$       &$\pi^+\pi^+\pi^-X$               &$0.103 \pm0.003$   &\cite{pdg2020}&Scaled by ${\mathcal B}(\eta\to \pi^+\pi^-X) \sim 27.4\%$\\
Others                          &$0.532\pm0.034$       &$\pi^+\pi^+\pi^-X$               &$0.426\pm0.027$ &\cite{pdg2020,kaikai}&Dominated by $\pi^+\eta(\bar K^0\pi^0, \pi^0, \pi^0\pi^0, \pi^0\pi^0\pi^0)$ and $\pi^+\phi$~$\sim$ 80\%\\
\hline
Sum               &...   &...  &$14.74\pm0.53$ &...     &...\\
\hline
\hline
\end{tabular}
\end{center}
\end{table*}


\begin{thebibliography}{*}
\bibitem{LFUtest}
Y. Amhis {\it et al.} (Heavy Flavor Averaging Group),
\href{https://doi.org/10.1140/epjc/s10052-020-8156-7}{Eur. Phys. J. C 81, 226 (2021)};
Updated results available at \href{https://hflav-eos.web.cern.ch/hflav-eos/semi/spring21/html/RDsDsstar/RDRDs.html}
{https://hflav-eos.web.cern.ch/hflav-eos/semi/spring21/html/RDsDsstar/RDRDs.html}

\bibitem{LHCb1}
R. Aaij {\it et al.} (LHCb Collaboration),
\href{https://journals.aps.org/prl/abstract/10.1103/PhysRevLett.120.171802}{Phys. Rev. Lett. {\bf 120}, 171802 (2018).}

\bibitem{LHCb2}
R. Aaij {\it et al.} (LHCb Collaboration),
\href{https://journals.aps.org/prd/abstract/10.1103/PhysRevD.97.072013}{Phys. Rev. D {\bf 97}, 072013 (2018).}


\bibitem{bes3-Ds-3pix}
M. Ablikim {\it et al.} (BESIII Collaboration),
\href{https://arxiv.org/abs/2212.13072}{arXiv:2212.13072}.

\bibitem{pdg2020}
P. A. Zyla {\it et al.} (Particle Data Group), Prog. Theor. Exp. Phys. {\bf 2022}, 083C01 (2022).

\bibitem{wuxiao}
M. Ablikim {\it et al.} (BESIII Collaboration),
\href{https://journals.aps.org/prd/abstract/10.1103/PhysRevD.105.032009}{Phys. Rev. D {\bf 105}, 032009 (2022).}

\bibitem{kaikai}
M. Ablikim {\it et al.} (BESIII Collaboration),
\href{https://arxiv.org/abs/2206.13864}{arXiv:2206.13864}.

\bibitem{lanxing}
M. Ablikim {\it et al.} (BESIII Collaboration),
\href{https://arxiv.org/abs/2205.14031}{arXiv:2205.14031}.

\bibitem{lum_bes3}
M. Ablikim {\it et al.} (BESIII Collaboration),
\href{https://iopscience.iop.org/article/10.1088/1674-1137/37/12/123001}{Chin. Phys. C {\bf 37}, 123001 (2013);}
\href{https://doi.org/10.1016/j.physletb.2015.11.043}{Phys. Lett. B {\bf 753}, 629 (2016).}

\bibitem{BESIII}
M. Ablikim {\it et al.} (BESIII Collaboration),
\href{https://doi.org/10.1016/j.nima.2009.12.050}{Nucl. Instrum. Meth. A {\bf 614}, 345 (2010).}

\bibitem{Yu:IPAC2016-TUYA01}
C.~H.~Yu {\it et al.},
\href{doi:10.18429/JACoW-IPAC2016-TUYA01}{Proceedings of IPAC2016, Busan, Korea, 2016.}

\bibitem{dectector}
K.X. Huang, et al.,
\href{https://doi.org/10.1007/s41365-022-01133-8}{Nucl. Sci. Tech. 33, 142 (2022).}

\bibitem{geant4}
S. Agostinelli {\it et al.} (GEANT4 Collaboration),
\href{https://doi.org/10.1016/S0168-9002(03)01368-8}{Nucl. Instrum. Meth. A {\bf 506}, 250 (2003).}

\bibitem{kkmc}
S. Jadach, B. F. L. Ward, and Z. Was,
\href{https://linkinghub.elsevier.com/retrieve/pii/S0010465500000485}{Comp. Phys. Commu. {\bf 130}, 260 (2000);} \href{https://journals.aps.org/prd/abstract/10.1103/PhysRevD.63.113009}{Phys. Rev. D {\bf 63}, 113009 (2001).}

\bibitem{evtgen}
D.~J.~Lange,
\href{https://doi.org/10.1016/S0168-9002(01)00089-4} {Nucl. Instrum. Meth. A {\bf 462}, 152 (2001);}
R.~G.~Ping,
\href{https://doi.org/10.1088/1674-1137/32/8/001}{Chin. Phys. C {\bf 32}, 599 (2008).}

\bibitem{lundcharm}
J. C. Chen, G. S. Huang, X. R. Qi, D. H. Zhang, and Y. S. Zhu,
\href{https://journals.aps.org/prd/abstract/10.1103/PhysRevD.62.034003}{Phys. Rev. D {\bf 62}, 034003 (2000).}

\bibitem{lundcharm2}
R. L. Yang, R. G. Ping and H. Chen,
\href{http://cpl.iphy.ac.cn/10.1088/0256-307X/31/6/061301}{Chin. Phys. Lett {\bf 31}, 061301 (2014).}

\bibitem{photos}
E.~Richter-Was,
\href{https://doi.org/10.1016/0370-2693(93)90062-M`}{Phys. Lett. B {\bf 303}, 163 (1993).}

\bibitem{MARKIII1}
R. M. Baltrusaitis {\it et al.} (Mark III Collaboration),
\href{https://doi.org/10.1103/PhysRevLett.56.2140}{Phys.Rev. Lett. 56, 2140 (1986).}

\bibitem{MARKIII2}
J. Adler {\it et al.} (Mark III Collaboration),
\href{https://doi.org/10.1103/PhysRevLett.60.89}{Phys. Rev. Lett. 60, 89 (1988).}

\bibitem{refcp3}
T.Gershon, J. Libby, and G. Wilkinson,
\href{https://doi.org/10.1016/j.physletb.2015.08.063}{Phys. Lett. B {\bf 750}, 338 (2015).}

\bibitem{refcp4}
T. Evans {\it et al.},
\href{https://doi.org/10.1016/j.physletb.2016.04.037}{Phys. Lett. B {\bf 757}, 520 (2016);}
\href{https://doi.org/10.1016/j.physletb.2016.11.021}{{\bf 765}, 402(E) (2017).}

\bibitem{A1}
Heavy Flavor Averaging Group (HFLAV),
\href{http://www.slac.stanford.edu/xorg/hflav/charm/}{(http://www.slac.stanford.edu/xorg/hflav/charm/).}

\bibitem{refcp1}
T. Evans {\it et al.},
\href{https://doi.org/10.1016/j.physletb.2016.04.037}{Phys. Lett. B {\bf 757}, 520 (2016);}
\href{https://doi.org/10.1016/j.physletb.2016.11.021}{{\bf 765}, 402(E) (2017).}

\bibitem{refcp2}
M. Ablikim {\it et al.} (BESIII Collaboration),
\href{https://journals.aps.org/prd/abstract/10.1103/PhysRevD.100.072006}{Phys. Rev. D {\bf 100}, 072006 (2019).}

\bibitem{bes3-pimuv}
M. Ablikim {\it et al.} (BESIII Collaboration),
\href{https://journals.aps.org/prl/abstract/10.1103/PhysRevLett.121.171803}{Phys. Rev. Lett. {\bf 121}, 171803 (2018).}

\bibitem{bes3-etaX}
M.~Ablikim {\it et al.} (BESIII Collaboration),
\href{https://dx.doi.org/10.1103/PhysRevLett.124.241803}{Phys. Rev. Lett. {\bf 124}, 241803 (2020).}

\bibitem{ARGUS}
H. Albrecht {\it et al.} (ARGUS Collaboration),
\href{https://doi.org/10.1016/0370-2693(90)91293-K}{Phys. Lett. B {\bf 241}, 278 (1990).}

\bibitem{MC}
M. Lefebvre, R. K. Keeler, R. Sobie and J. White,
\href{https://linkinghub.elsevier.com/retrieve/pii/S0168900200003235}{Nucl. Instr. Meth. A 451 (2000) 520.}

\bibitem{bes3-klev}
M.~Ablikim {\it et al.} (BESIII Collaboration),
\href{https://journals.aps.org/prd/abstract/10.1103/PhysRevD.92.112008}{Phys. Rev. D. {\bf 92}, 112008 (2015).}

\end{thebibliography}
\end{document}